\documentclass[12pt,a4]{article}
\usepackage{relsize}
\usepackage{array}
\usepackage{a4wide}
\usepackage[hidelinks]{hyperref}
\usepackage{latexsym}
\usepackage{cite}
\usepackage{comment}
\usepackage{caption}
\usepackage{mathrsfs}
\usepackage{tikz}
\usetikzlibrary{arrows.meta,calc,decorations.markings,math,arrows.meta,shapes.misc, decorations.pathreplacing}
\usepackage{amssymb}
\usepackage{amsmath}
\usepackage{graphicx}
\usepackage{enumerate}
\usepackage{todonotes}
\usepackage{a4wide}
\usepackage{tensor}
\usepackage{comment}
\usepackage{dsfont}
\usepackage{braket}

\usepackage{filecontents}

\newcommand{\be}{\begin{equation}\label}
\newcommand{\ee}{\end{equation}}
\newcommand{\bea}{\begin{eqnarray}\label}
\newcommand{\eea}{\end{eqnarray}}

\makeatletter
\newcommand*{\textoverline}[1]{$\overline{\hbox{#1}}\m@th$}
\newcommand*\bigcdot{\mathpalette\bigcdot@{.65}}
\newcommand*\bigcdot@[2]{\mathbin{\vcenter{\hbox{\scalebox{#2}{$\m@th#1\bullet$}}}}}
\makeatother

\newcommand{\nn}{\nonumber}

\newcommand{\Op}{\mathcal{O}}

\numberwithin{equation}{section} 

\begin{document}

 \centerline{\LARGE \bf {\sc Non-Lorentzian $SU(1,n)$ Spacetime Symmetry    } } 
 \medskip
 
  \centerline{\LARGE \bf {\sc in Various Dimensions  } } \vspace{1cm}

  \centerline{
   {\large {\bf  {\sc N.~Lambert,${}^{\,a}$}}\footnote{E-mail address: \href{neil.lambert@kcl.ac.uk}{\tt neil.lambert@kcl.ac.uk}}     \,  {\sc R.~Mouland${}^{\,b}$}\footnote{E-mail address: \href{r.mouland@damtp.cam.ac.uk}{\tt r.mouland@damtp.cam.ac.uk}} \, {\sc and T.~Orchard${}^{\,a}$}\footnote{E-mail address: \href{tristan.orchard@kcl.ac.uk}{\tt tristan.orchard@kcl.ac.uk}}   }}  
     
\vspace{1cm}
\centerline{${}^a${\it Department of Mathematics}}
\centerline{{\it King's College London }} 
\centerline{{\it The Strand, WC2R 2LS, UK}} 
  
\vspace{1cm}
\centerline{${}^b${\it Department of Applied Mathematics and Theoretical Physics}}
\centerline{{\it University of Cambridge}} 
\centerline{{\it  Cambridge, CB3 0WA, UK}}

\vspace{1.0truecm}

 
\thispagestyle{empty}

\centerline{\sc Abstract}\

\noindent We discuss non-Lorentzian    Lagrangian field theories in $2n-1$  dimensions that admit an $SU(1,n)$ spacetime symmetry which includes a scaling transformation. These can be obtained by a conformal compactification of a $2n$-dimensional Minkowskian conformal field theory. We discuss the symmetry algebra, its representations including primary fields and unitarity bounds. We also give various examples of free theories in a variety of dimensions and a discussion of how to reconstruct the parent $2n$-dimensional theory.

\newpage

\section{Introduction}
 
Lorentz symmetry plays a crucial role in many applications of Quantum Field Theory but it is not necessary. Indeed the condensed matter community more often than not looks at theories without it. This opens the door to additional spacetime symmetries such as the Bargmann, Carroll and Schr\"odinger groups. In particular non-Lorentzian conformal field theories have now received considerable attention and reveal many interesting features, for example see \cite{Nishida:2007pj,Nishida:2010tm,Goldberger:2014hca,Karananas:2021bqw,Doroud:2016mfv}.

It is well-known that one way  to construct non-Lorentzian theories with Schr\"odinger symmetry is to  reduce a Lorentzian theory of one higher dimension on a null direction.  From the higher dimensional perspective such null reductions are somewhat unphysical but that need not concern us if we are only interested in the features of the reduced theory. Indeed the (null) Kaluza-Klein momentum is often associated with particle number and, in contrast to traditional Kaluza-Klein theories, one need not truncate the action to the zero-modes but rather any given Fourier mode. The resulting theories are interesting in themselves and have applications in Condensed Matter Systems and  DLCQ constructions (where one does have to try to make sense of a null reduction). 

Here we will explore theories with novel spacetime $SU(1,n)$ symmetry. These can be obtained by reducing a Lorentzian conformal field theory (CFT) along a null direction in conformally compactified Minkowski space.  A key novelty here is that the conformal null reduction can be inverted so that the non-compact higher dimensional theory can in principle be reconstructed from the reduced theory provided all Kaluza-Klein modes are retained. The effect of such a reduction is to induce an $\Omega$-deformation into the reduced theory. Since the $SU(1,n)$ symmetry acts separately on each Fourier mode we can truncate our actions to any given Fourier mode number. Or we can keep them all and reconstruct the original theory.  

 We also comment that our interest in these models has arisen through an explicit  class of supersymmetric non-Abelian gauge theories in five-dimensions with $\frak{su}(1,3)$ symmetry \cite{Lambert:2019jwi,Lambert:2019fne,Lambert:2020jjm}. In particular, in these models the role of the Kaluza-Klein momentum $P_+$ is played by the instanton number leading to an additional $\frak{u}(1)$ symmetry.  The realisation of the full non-Abelian six-dimensional $(2,0)$ theory is proposed to arise through the inclusion of instanton operators \cite{Lambert:2020zdc,Lambert:2021mnu,Lambert:2021fsl}.

In this paper we wish to illustrate some of general aspects of such theories. 
In Section 2 we will outline a construction  by dimensional reduction of a CFT on conformally compactified Minkowski space and give the corresponding AdS interpretation. In Section 3 we discuss  various properties  of the $SU(1,n)$ symmetry algebra such as primary fields,  unitarity bounds and  its relation to conventional non-relativistic conformal symmetry. In Section 4 we  discuss a superconformal extension that is possible in the case of five-dimensions and construct some BPS bounds. In section 5 we will give explicit examples of theories with $SU(1,n)$ symmetry. In the interest of simplicity we will only consider free theories here, although, as mentioned above,  interacting theories can be constructed. In section 6 we will outline how, by retaining the entire Kaluza-Klein tower of fields, one can reconstruct the 2-point functions of of the parent $2n$-dimensional theory. Finally in section 7 we give our conclusions and comments.

\section{Construction Via Conformal Compactification}

We start with $2n$-dimensional Minkowski spacetime in lightcone coordinates with metric 
\begin{align}
	ds^2_M = \eta_{\mu\nu}d\hat x^\mu d\hat x^\nu =  -2d\hat x^+d\hat x^- + d\hat x^id \hat x^i\ ,
	\label{eq: 6d Minkowski metric}
\end{align}
where $\mu\in\{+,-,i\}$, $i=1,2,...,2n-2$, and perform the coordinate transformation\footnote{It is curious to note that this transformation is similar to the transformation used in \cite{Goldberger:2014hca} to convert to the so-called oscillator frame, along with an $x^+$-dependent rotation by $\Omega_{ij}$.}
\begin{align}\label{eq: coordinate transformation}
\hat x^+ &= 2R \tan (x^+/2R) \, , \nonumber\\ 
\hat x^- &= x^- +\frac{1}{4R}x^ix^i\tan (x^+/2R)\, ,\nonumber\\   
\hat x^i &=\frac{\cos(x^+/2R)x^i - \sin(x^+/2R) R\Omega_{ij}x^j	}{\cos(x^+/2R)}\ .
\end{align}
Here $\Omega_{ij}$ is a constant anti-symmetric  matrix that satisfies 
\begin{align}
\Omega_{ij}\Omega_{jk} = -R^{-2}\delta_{ik}	\, .
\end{align}
Note, we can always perform a rotation in the $x^i$ directions so as to bring $\Omega_{ij}$ to a canonical form; in particular, one can always find orthogonal matrix $M$ such that
\begin{align}
\Omega\to   M\Omega M^{-1} = M \Omega M^T = \frac{1}{R}\begin{pmatrix}
  	0 & \mathds{1}_{n-1}		\\
  	-\mathds{1}_{n-1} & 0
  \end{pmatrix}\ .
  \label{eq: orthog transformation}
\end{align}
This coordinate transformation leads to the  metric
\begin{align}\label{5dmink}
ds^2_M = \hat g_{\mu\nu}d x^\mu d x^\nu =\frac{-2dx^+(dx^- + \frac12 \Omega_{ij}x^jdx^i) + dx^idx^i}{\cos^2(x^+/2R)}\ .	
 \end{align}
Following this we perform a Weyl transformation $ds^2_\Omega = {\cos^2(x^+/2R)}ds^2_M$ to find
\begin{align}
ds^2_\Omega = g_{\mu\nu}x^\mu x^\nu =  -2dx^+\left(dx^- + \frac12 \Omega_{ij}x^jdx^i\right) + dx^idx^i \ .
\label{eq: conformally compactified metric}	
\end{align} 
Under such a conformal transformation a scalar operator $\hat {\cal O}(\hat x)$ of dimension $\hat \Delta$ is mapped to the operator ${\cal O}(x)$ by
\begin{align}
\hat {\cal O}(\hat x)	=\cos^{\hat \Delta}(x^+/2R){\cal O}(x)\ .
\end{align}
Note the range of $x^+\in (-\pi R,\pi R)$ is finite. Thus we can  conformally compactify the $x^+$ direction of $2n$-dimensional Minkowski space by $x^+\in [-\pi R,\pi R]$. In which case we can write ${\cal O}(x)$ in a Fourier expansion:
\begin{align}
{\cal O}(x) = \sum_k e^{ikx^+/R} {\cal O}^{(k)}(x)	\ ,
\end{align}
where for now we keep the range of $k$ general, {\it e.g.} integer or half-integer.
Lastly is helpful to note that the metric and inverse metric are
\begin{align}
g_{\mu\nu} =\left(\begin{matrix}
		0 & -1 & -\frac12\Omega_{ik}x^k  \\
		-1& 0& 0\\
		-\frac12\Omega_{ik}x^k  &0&\delta_{ij} 
	\end{matrix}\right)\qquad 	g^{\mu\nu} =\left(\begin{matrix}
		0 & -1 & 0\\
		-1& |x^2|/4R^2 & -\frac12 \Omega_{jk}x^k\\
		0&-\frac12\Omega_{ik}x^k &\delta_{ij} 
	\end{matrix}\right)\ .
\end{align}

\subsection{Dual AdS slicing}\label{subsec: AdS}

As we have seen, the metric $ds_\Omega^2$ in (\ref{eq: conformally compactified metric}) is conformal to $2n$-dimensional Minkowski space, and hence can be realised as the conformal boundary of Lorentzian $\text{AdS}_{2n+1}$. Indeed, a particular slicing of $\text{AdS}_{2n+1}$ that makes this form for the conformal boundary manifest has long been known in the literature \cite{Pope:1999xg}. Let us review this construction now, with a focus using this holographic perspective to probe the form of the conformal algebra on the boundary.

Let $Z^a$, $a=,0,1,\dots,n$ be a set of $(n+1)$ complex coordinates, and $\eta_{ab}=\text{diag}(-1,1,\dots,1)$. Then, when constrained to
\begin{align}
  \eta_{ab}Z^a \bar{Z}^b = -1\ ,
  \label{eq: Z constraint}
\end{align}
the $Z^a$ provide coordinates on Lorentzian $\text{AdS}_{2n+1}$, with metric given by
\begin{align}
  ds^2 = \eta_{ab}dZ^ad\bar{Z}^b\ ,
  \label{eq: AdS metric}
\end{align}
suitably pulled back to solutions of (\ref{eq: Z constraint}).

Next, we can parameterise solutions to the constraint (\ref{eq: Z constraint}) with $2n+1$ real coordinates $(y,x^+, x^-, x^i)$. We have\footnote{As our focus is on continuous conformal symmetries on the boundary, it is sufficient for our purposes to consider this a local parameterisation of $\text{AdS}_{2n+2}$, and thus neglect global features of this real coordinate choice.} 
\begin{align}
  Z^0		&=	e^{ix^+/2R} \left(\cosh\left(\frac{y}{2}\right)+ \frac{1}{2} e^{y/2}\left(iR x^- + \tfrac{1}{4}|\vec{x}|^2\right)\right)				\nn\\
  Z^n		&= 	e^{ix^+/2R} \left(\sinh\left(\frac{y}{2}\right)- \frac{1}{2} e^{y/2}\left(iR x^- + \tfrac{1}{4}|\vec{x}|^2\right)\right)					\nn\\
  Z^A		&= \frac{1}{2}e^{ix^+/2R}e^{y/2} \left(M(\vec{x}+ iR\Omega \vec{x})\right)^A ,\quad A=1,\dots,n-1\ ,
\end{align}
where here $M$ is the orthogonal matrix appearing in (\ref{eq: orthog transformation}).

These coordinates provide a description for $\text{AdS}_{2n+1}$ as a one-dimensional fibration over a non-compact form of $n$-dimensional complex projective space, sometimes denoted $\tilde{\mathbb{CP}}^n$. In this construction, $x^+$ is the coordinate along the fibre. The metric (\ref{eq: AdS metric}) now takes the form
\begin{align}
  ds^2 = - \frac{1}{4R^2}\left(dx^+ + R^2e^y \left(dx^- + \frac{1}{2}\Omega_{ij}x^j dx^i\right)\right)^2 + ds_{\widetilde{\mathbb{CP}}^n}\ ,
\end{align}
where
\begin{align}
  ds_{\widetilde{\mathbb{CP}}^n} = \frac{1}{4}dy^2 + \frac{1}{4}e^y dx^i dx^i + \frac{R^2}{4}e^{2y}\left(dx^- + \frac{1}{2}\Omega_{ij}x^j dx^i\right)^2\ ,
\end{align}
can be identified as the metric on $\widetilde{\mathbb{CP}}^n$, with isometry group $SU(1,n)$. Then, projecting orthogonally to the orbits of $\partial/\partial x^+$, we land precisely on $\widetilde{\mathbb{CP}}^n$ with metric $ds_{\widetilde{\mathbb{CP}}^n}$, as claimed. 

To go to the conformal boundary, we now restrict to a surface of constant $y$, and take $y$ large. It is then clear that as we do so, the metric approaches the form
\begin{align}
  ds^2 \to \frac{1}{4}e^y \left(-2dx^+\left(dx^- + \frac12 \Omega_{ij}x^jdx^i\right) + dx^idx^i \right)\ ,
\end{align}
thus recovering the form of the metric $ds_{\Omega}^2$.

Finally, let us discuss symmetries. Each isometry in the bulk, described by some Killing vector field, corresponds to a conformal symmetry on the boundary. The full set of such symmetries form the algebra $\frak{so}(2,2n)$.
It will be useful for what follows, however, to identify the subalgebra of boundary conformal symmetries that commute with translations along the $x^+$ direction. We see that this subalgebra can be identified with the subalgebra of bulk isometries that commute with translations along the fibre. It is hence given by $\frak{u}(1)\oplus \frak{su}(1,n)$, where $\frak{u}(1)$ describes translation along the fibre, while $\frak{su}(1,n)$ forms the algebra of isometries of the $\widetilde{\mathbb{CP}}^n$ transverse to the fibres.

\subsection{Symmetries under dimensional reduction}
\label{subsec:reduc}

Each continuous spacetime symmetry of a conformal field theory on Minkowski space is generated by an operator $G$, with the set of all such operators forming the algebra $\frak{so}(2,2n)$ under commutation. We take the conventional basis, made up of translations $\hat{P}_\mu$, Lorentz transformations $\hat{M}_{\mu\nu}$, dilatation $\hat{D}$ and special conformal transformations $\hat{K}_\mu$.
  
Each operator $G$ in turn correspond to a conformal Killing vector $G_\partial$ of the metric $ds^2_M$. Explicitly, these are
\begin{align}
  i (\hat{P}_\mu)_\partial &= \hat{\partial}_\mu																	& \hat{\omega} &= 0		\nn\\
  i(\hat{M}_{\mu\nu})_\partial &= \hat{x}_\mu \hat{\partial}_\nu	 - \hat{x}_\nu \hat{\partial}_\mu	  			& \hat{\omega} &= 0						\nn\\
  i(\hat{D})_\partial &= \hat{x}^\mu\hat{\partial}_\mu															& \hat{\omega} &= 1						\nn\\
  i(\hat{K}_\mu)_\partial &= \hat{x}_\nu \hat{x}^\nu \hat{\partial}_\mu - 2 \hat{x}_\mu \hat{x}^\nu \hat{\partial}_\nu	& \hat{\omega} &= -2\hat{x}_\mu\ ,
  \label{eq: so(2,n) generators}
\end{align}
with indices raised and lowered with the Minkowski metric $\eta_{\mu\nu}$. Each of these vector fields $G_\partial$ then satisfies $\mathcal{L}_{i G_\partial} \eta = 2\hat{\omega}\eta$, where $\mathcal{L}_{iG_\partial}$ is the Lie derivative along $iG_{\partial}$, and the Weyl factors $\hat{\omega}$ are as given.

Their non-vanishing commutators are 
\begin{align}
  	i[\hat{M}_{\mu\nu},\hat{P}_\rho]		&=	\eta_{\nu\rho}\hat{P}_\mu - \eta_{\mu\rho}\hat{P}_\nu	& i[\hat{P}_\mu,\hat{D}]		&=		\hat{P}_\mu			\nn\\
  	i[\hat{M}_{\mu\nu},\hat{K}_\rho]		&=	\eta_{\nu\rho}\hat{K}_\mu - \eta_{\mu\rho}\hat{K}_\nu		& i[\hat{K}_\mu,\hat{D}]		&=		-\hat{K}_\mu 		\nn\\
  	i[\hat{M}_{\mu\nu},\hat{M}_{\rho\sigma}]		&=	\eta_{\nu\rho}\hat{M}_{\mu\sigma}+ \eta_{\mu\sigma}\hat{M}_{\nu\rho} - 	\eta_{\mu\rho}\hat{M}_{\nu\sigma} - \eta_{\nu\sigma}\hat{M}_{\mu\rho}	& i[\hat{P}_\mu,\hat{K}_\nu]	&=	2(\hat{M}_{\mu\nu}-\eta_{\mu\nu}\hat{D})\ .
\end{align}
We can then perform the coordinate transformation (\ref{eq: coordinate transformation}) followed by the Weyl rescaling to arrive at the metric $g$ as in (\ref{eq: conformally compactified metric}). The operators $\{\hat{P}_\mu, \hat{M}_{\mu\nu},\hat{D},\hat{K}_\mu\}$ still generate the theory's spacetime symmetries, and the corresponding vector fields are also conformal Killing vectors of the metric $g$, albeit with shifted Weyl factors given by $\omega= \hat{\omega} - \frac{1}{2R}\tan\left(\frac{x^+}{2R}\right)iG_\partial^+$. Then, for each vector field, $\mathcal{L}_{iG_\partial} g = 2\omega g$.

It is now straightforward to see that translations along $x^+$ are an isometry\footnote{ Translations in $x^+$ are a conformal symmetry of the original metric $\hat{g}$, with non-trivial Weyl factor, which is cancelled when we perform the Weyl rescaling to arrive at $g$} of the metric $g$. In terms of the original Minkowski symmetry generators, this is realised by the combination
\begin{align}
  P_+:= \hat{P}_+ + \frac{1}{4}\Omega_{ij}\hat{M}_{ij} + \frac{1}{8R^2}\hat{K}_- \quad \longrightarrow \quad i(P_+)_\partial = \partial_+\ .
\end{align}
Then, given some conformal field theory on $2n$-dimensional Minkowski space, we can perform a Kaluza-Klein on the $x^+$ interval. At the level of the symmetry algebra, this amounts to choosing a basis for the space of local operators which diagonalises $P_+$. The resulting operators are Fourier modes on the $x^+$ interval. They fall into representations of the centraliser of $P_+$ within $\frak{so}(2,2n)$, which we call $\frak{h}$.

A basis for the subalgebra $\frak{h}$ in terms of the generators (\ref{eq: so(2,n) generators}) is given by
 \begin{align}
  	P_+ 	&= \hat{P}_+ + \tfrac{1}{4}\Omega_{ij}\hat{M}_{ij} + \tfrac{1}{8R^2}\hat{K}_-	
  			&&\longrightarrow\quad i(P_+)_\partial = \partial_+ 	\nn\\
  	H 		&= \hat{P}_-		
  			&&\longrightarrow\quad i(H)_\partial = \partial_-	\nn\\
  	P_i		&=	\hat{P}_i +\tfrac{1}{2}\Omega_{ij} \hat{M}_{j-}
  			&&\longrightarrow\quad i(P_i)_\partial= \tfrac{1}{2}\Omega_{ij} x^j \partial_- + \partial_i		\nn\\
  	B 		&=	\tfrac{1}{2} R\, \Omega_{ij} \hat{M}_{ij}
  			&&\longrightarrow\quad i(B)_\partial=  R \,\Omega_{ij}x^i \partial_j										\nn\\
  	J^\alpha &=	\tfrac{1}{2} L^\alpha_{ij} \hat{M}_{ij}
  			&&\longrightarrow\quad i(J^\alpha)_\partial= L^\alpha_{ij}x^i \partial_j										\nn\\
  	T		&=	\hat{D} - \hat{M}_{+-}
  			&&\longrightarrow\quad i(T)_\partial= 2x^- \partial_- + x^i \partial_i							\nn\\
  	G_i	&=	\hat{M}_{i+} - \tfrac{1}{4}\Omega_{ij} \hat{K}_j
  			&&\longrightarrow\quad i(G_i)_\partial=  x^i \partial_+ + \left( \tfrac{1}{2}\Omega_{ij} x^- x^j - \tfrac{1}{8}R^{-2} x^j x^j x^i \right)\partial_- + x^- \partial_i  				\nn\\
			&&&  \hspace{35mm} +\tfrac{1}{4}( 2\Omega_{ik}x^k x^j + 2\Omega_{jk}x^k x^i - \Omega_{ij}x^k x^k )\partial_j				\nn\\
  	K	&=	\tfrac{1}{2}\hat{K}_+
  			&&\longrightarrow\quad i(K)_\partial=  x^i x^i \partial_+ + ( 2 ( x^- )^2 - \tfrac{1}{8} R^{-2} ( x^i x^i )^2 )\partial_-															\nn\\			
			&&&  \hspace{35mm} +( \tfrac{1}{2} \Omega_{ij} x^j x^k x^k + 2 x^- x^i )\partial_i\ ,
			\label{eq: new gens in terms of old}
\end{align}
where the $J^\alpha$ are absent for $n=1,2$, and otherwise $\alpha=1,\dots, n^2 - 2n$. Here, the $L^\alpha_{ij}$ are constant matrices that are explained below.

Then, these vector fields are indeed conformal Killing vector fields of $g$, as $\mathcal{L}_{iG_\partial } g = 2\omega g$ with Weyl factor given by
\begin{align}
  T:\quad 		&\omega=1									\nn\\
  G_{i}:\quad 	&\omega=\tfrac{1}{2}\Omega_{ij}x^j	\nn\\
  K:\quad		&\omega= x^-\ ,
\end{align}
and vanishing for the other generators.

Let us identify the subalgebra of pure rotations within $\frak{h}$, and in particular identify the matrices $L^\alpha$. First, the case of $n=1$ is somewhat trivial, as we have no spatial directions, and so no rotations to start with. Similarly straightforward is $n=2$, whereby we can simply take $B=\frac{1}{2}R\,\Omega_{ij}\hat{M}_{ij}$ as the single generator of the rotation subalgebra $\frak{so}(2)$, which is easily seen the commute with $P_+$ and hence survive the reduction.

So let us take $n\ge 3$. We may, a priori, consider a general spatial rotation of the form $A_{ij}\hat{M}_{ij}$ for any $(2n-2)\times (2n-2)$ matrix $A$ with $A_{ij} = - A_{ji} $, forming $\frak{so}(2n-2)\subset \frak{so}(2,2n)$. Then, it is clear from the form of $P_+$ that this rotation commutes with $P_+$ and thus lies within $\frak{h}$ precisely if $[A,\Omega]_{ij} = A_{ik}\Omega_{kj} - \Omega_{ik} A_{kj} = 0$. It then follows from the relation (\ref{eq: orthog transformation}) that $A$ must be similar to an element of $\frak{so}(2n-2)\,\cap\, \frak{sp}(2n-2) \cong \frak{u}(1)\oplus \frak{su}(n-1)$. Hence, the set of matrices $A$ form a $(2n-2)$-dimensional representation of $\frak{u}(1)\oplus \frak{su}(n-1)$. In particular, we can take $\Omega$ itself to span the $\frak{u}(1)$ factor, while we write $L^\alpha$, $\alpha=1,\dots, n^2-2n$ form the generators of $\frak{su}(n-1)$. We have $[L^\alpha,L^\beta]_{ij} = f^{\alpha\beta}{}_{\gamma}L^\gamma_{ij}$ for $f^{\alpha\beta}{}_{\gamma}$ the structure constants of $\frak{su}(n-1)$. Note, by construction, the matrices $(\Omega L^\alpha)_{ij}=\Omega_{ik} L^\alpha_{kj}$ are symmetric and traceless for each $\alpha$. 

Thus, for $n\ge 3$ the total rotation subalgebra is $\frak{u}(1)\oplus \frak{su}(n-1) \subset \frak{h}$, spanned by $\{B, J^\alpha\}$. For example, for the first non-trivial case $n=3$, one can show from the relation (\ref{eq: orthog transformation}) that all choices of the $4\times 4$ matrices $\Omega_{ij}$ fall into two classes: those that are anti-self-dual, and those that are self-dual. These two cases correspond to $\det(M)=+1$ and $\det(M)=-1$, respectively. Then, for anti-self-dual (self-dual) $\Omega_{ij}$, one can choose for their $L^\alpha_{ij}$ the self-dual (anti-self-dual) 't Hooft matrices. 

Finally, let us state the commutation relations for the algebra $\frak{h}$. The commutators of the rotation subalgebra $\text{span}\{B, J^\alpha\}$ both amongst themselves and with the other generators can be summarised as follows. As we have seen, $\{B,J^\alpha\}$ form a basis for $\frak{u}(1)\oplus \frak{su}(n-1)$. The commutation relations descend from those of the $L^\alpha$, so that
\begin{align}
  i[B, J^\alpha]	= 0 \qquad i[J^\alpha,J^\beta] = f^{\alpha\beta}{}_\gamma J^\gamma\ .
\end{align}
The remaining generators are sorted into `scalar' generators $\mathcal{S}=\{P_+, H,  T, K\}$ which are inert under rotations, $i[B, \mathcal{S}]=0, i[J^\alpha, \mathcal{S}]=0$, and otherwise `vector' generators $\mathcal{V}_i =\{P_i, G_i\}$ which transform as
\begin{align}
  i[B, \mathcal{V}_i] = - R\,\Omega_{ij} \mathcal{V}_j,\qquad i[J^\alpha, \mathcal{V}_i] = - L^\alpha_{ij} \mathcal{V}_j\ ,
\end{align}
All remaining commutators are found to be
\begin{align}
		i[G_i,P_j] \ &= \ -\delta_{ij} P_+ - \tfrac{1}{2}\Omega_{ij} T +\tfrac{1}{2R}\tfrac{n+1}{n-1} \delta_{ij} B + \beta^{\alpha}_{ij} J^\alpha    \, ,	
	&	i[T,H] \ 	&= \ -2 H					\, ,\nn\\
  		i[T,K] \	 &= \ 2K \, ,					
  	&	i[H,P_i] 				\ &= \ 0		  				\, ,\nn\\
  		i[K,H] \ &= \ -T	\, ,					  		
  	&	i[H,G_{i}] 		\ &= \ P_i   			\, ,\nn\\
		i[G_{i},G_{j}] \ &= \ -  \Omega_{ij} K \, , 
	&	i[K,P_i] 				\ &= \ -G_{i} 		\, ,\nn\\
  		i[T,P_i] 	\ &= \ -P_i \, ,		
  	&	i[K,G_{i}] 			\ &= \  0  						\, ,\nn\\
 		i[T,G_{i}] \ &= \ G_{i} \, ,
 	&	i[P_i,P_j] 		\ &= \ -\Omega_{ij} H						\, .
 	\label{eq: algebra}
\end{align}
where the coefficient in front of $B$ in the commutator $i[G_i,P_j]$ holds down to $n=2$. Further, we denote by $\beta^\alpha_{ij}$ the constants such that
\begin{align}
   \frac{1}{2}\left(\delta_{jk} \Omega_{il} + \delta_{ik} \Omega_{jl} - \delta_{jl} \Omega_{ik}-\delta_{il} \Omega_{jk} - \frac{2}{n-1}\delta_{ij} \Omega_{kl}\right) = \beta^\alpha_{ij} L^\alpha_{kl}\ .
\end{align}
One can show that this equation can always be uniquely solved for the $\beta^\alpha_{ij}$, for any choice of the basis $L^\alpha_{ij}$ for $\frak{su}(n-1)$, and further that $\beta^\alpha_{ij}=\beta^\alpha_{ji}$ and $\beta^\alpha_{ii}=0$.

Following the discussion in Section \ref{subsec: AdS}, we identify $\frak{h} = \frak{u}(1) \oplus \frak{su}(1,n)$. This splitting is made explicit by adjusting the rotation $B$ to twist along the $x^+$ interval. In detail, we define
\begin{align}
  \tilde{B} := B - 2R \left(\frac{n-1}{n+1}\right)P_+\ .
\end{align}
Then, $\{H, P_i, \tilde{B}, J^\alpha, T, G_i, K\}$ form a basis for $\frak{su}(1,n)$, while $P_+$ generates the $\frak{u}(1)$ factor.

\section{Primary Operators and Their Properties}\label{subsec: primaries}

So let us now consider a $(2n-1)$-dimensional theory with $SU(1,n)$ symmetry. Given some operator $\Phi(0)$ at the origin $(x^-, x^i) = (0,0)$, we say it has scaling dimension $\Delta$ if it satisfies $[T,\Phi(0)]=i\Delta \Phi(0)$. Then, in direct analogy with the Schr\"odinger algebra of conventional non-relativistic conformal field theory, we can straightforwardly construct further states also with definite charge under $T$.

We find that $\{H, K\}$ raise and lower scaling dimension by 2 units, respectively, so that if $\Phi(0)$ has scaling dimension $\Delta$, then $[H,\Phi(0)]$ has scaling dimension $(\Delta+2)$, while $[K,\Phi(0)]$ has $(\Delta-2)$. We have then also the pair $\{P_i, G_i\}$, which raise and lower scalig dimension by 1 unit, respectively.

Going further, we can generalise results from the $n=3$ case \cite{Lambert:2020zdc}, and define a primary operator at the origin $(x^-, x^i) = (0,0)$ by its transformation under the stabiliser of the origin within $\frak{u}(1)\oplus \frak{su}(1,n)$, generated by $\{P_+,B, J^\alpha, T, G_i, K\}$. We have,
\begin{align}
   [P_+, 	\Op(0)] 	&=		p_+ \Op(0) 				\nn\\
   [B,		\Op(0)	]	&=		-r_\Op[B]\Op(0)			\nn\\
   [J^\alpha, \Op(0)]	&=		-r_\Op[J^\alpha]	\Op(0)	\nn\\
   [T, \Op(0)]			&=		i\Delta \Op(0)			\nn\\
   [G_i, \Op(0)]		&=		0						\nn\\
   [K, \Op(0)]			&=		0\ .
   \label{eq: primary at origin}
\end{align}
Here, $r_\Op[B]\in \frac{1}{2}\mathbb{Z}$ denotes the charge of $\Op(0)$ under the rotation generated by $B$, while $r_\Op[J^\alpha]$ is a constant matrix acting on some unwritten discrete indices of $\Op(0)$, and forming an irreducible representation of the $\frak{su}(n-1)$ spanned by the $J^\alpha$, so that $[r_\Op[J^\alpha],r_\Op[J^\beta]] = r_\Op[[J^\alpha, J^\beta]]$. Finally, $p_+$ is the charge of $\Op(0)$ under $P_+$. It is clear that in any $(2n-1)$-dimensional theory found from a $2n$-dimensional CFT, we must have $p_+\in \frac{1}{R}\mathbb{Z}$, however one may in principle consider a broader class of theories not admitting a $2n$-dimensional interpretation, and thus without such a discreteness condition.

The key property of such a primary is that it is annihilated by the lowering operators $\{K, G_i\}$, and thus sits at the bottom of a tower of states generated by the raising operators $\{H, P_i\}$, known as usual as \textit{descendants}.

Given any operator $\Phi(0)$ at the origin, an operator at some point $(x^-, x^i)$ is defined by
\begin{align}
  \Phi(x) = \exp (-i \left(x^- H + x^i P_i\right)) \Phi(0) \exp (i \left(x^- H + x^i P_i\right))\ .
\end{align}
Then, requiring that at any point we have $\Phi(x+\epsilon)- \Phi(x) = \epsilon^- \partial_- \Phi(x) + \epsilon^i \partial_i \Phi(x)$ fixes the action of $H,P_i$ on $\Phi(x)$ \cite{Lambert:2020zdc}. Note, this is a somewhat more subtle computation than is encountered in relativistic conformal field theory, since the translation subalgebra $\text{span}\{H,P_i\}$ is non-Abelian.

One can in particular apply the transformation rules (\ref{eq: primary at origin}) along with the algebra (\ref{eq: algebra}) to determine the transformation properties of a primary $\Op(x)$ at a generic point. This generalisation of the known form for $n=3$ is left as an exercise for the reader.

\subsection{Recovering conventional non-relativistic conformal field theory}\label{subsec: DLCQ}

At the level of symmetries, the presence of conformal symmetry in the relativistic theory manifests itself as an enhancement of the Poincar\'{e} algebra to the conformal algebra. The analogous statement in non-relativistic theories is an enhancement of the inhomogeneous Galilean algebra---or rather, its central extension, the Bargmann algebra---to the Schr\"{o}dinger algebra. Let us denote by $\text{Schr}(d)$ the Schr\"odinger algebra governing the non-relativsitic conformal dynamics of a particle in $d$ spatial dimensions.

Then, $\text{Schr}(d)$ is realised precisely as the centraliser of a null translation within the conformal algebra $\frak{so}(2,d+2)$ of $\mathbb{R}^{1,d+1}$. The single central element of $\text{Schr}(d)$, often interpreted as particle number, is realised by this null translation.

Recall, we defined the subalgebra $\frak{h}=\frak{u}(1)\oplus \frak{su}(1,n)\subset \frak{so}(2,2n)$ as the centraliser of the generator $P_+$. It is clear that in the limit that $R\to\infty$, the coordinate transformation (\ref{eq: coordinate transformation}) and subsequent Weyl rescaling become trivial, and as such $P_+$ degenerates to become simply a null translation. Indeed, this is also evident from the form of $P_+$ in terms of the conventional conformal generators, as in (\ref{eq: new gens in terms of old}), where we see that as $R\to\infty$, we have $P_+\to \hat{P}_+$.

Hence, in the limit $R\to\infty$, the subalgebra $\frak{h}$ must become some subalgebra of $\text{Schr}(2n-2)\subset \frak{so}(2,2n)$. Note that $\frak{h}$ needn't give us the \textit{whole} Schr\"odinger algebra, since there may be elements within $\frak{so}(2,2n)$ that only commute with $P_+$ strictly in the $R\to\infty$ limit. Indeed, this is precisely what happens. It is evident that strictly in the $R\to\infty$ limit, the spatial 2-form $\Omega_{ij}$ drops out entirely, and thus the breaking of the rotation subalgebra $\frak{so}(2n-2)\to \frak{u}(1)\oplus \frak{su}(n-1)$ does not occur. One can indeed show that in taking the limit $R\to\infty$ and adding back in by hand the rotations broken by $\Omega_{ij}$ at finite $R$, we do indeed recover the Schr\"odinger algebra $\text{Schr}(2n-2)$.

Things therefore work smoothly at the level of the algebra. However, given a theory admitting the $\Omega$-deformed non-relativistic conformal symmetry $\frak{u}(1)\oplus \frak{su}(1,n)$, there is an additional step we should take in order to recover the correct global Schr\"odinger group. In particular, a particle interpretation requires that the particle number $N$ has discrete eigenvalues, corresponding in turn to a compactification of the null direction as $x^+\sim x^+ + 2\pi R_+$ for some $R_+$, which by a Lorentz boost is seen to be unphysical.

A convenient way to arrive at this setup---which from the $2n$-dimensional perspective coincides with that of DLCQ---is to first introduce an orbifold. In particular at finite $R$ the orbifold restricts to operators that are periodic but with period $2\pi R/K$ along the $x^+$ direction for some $K\in\mathbb{N}$. Equivalently, we project onto the Hilbert subspace of states with $P_+$ eigenvalue in $\frac{K}{R}\mathbb{Z}$. Now, taking $K,R\to\infty$ while holding their ratio $R_+:= R/K$ fixed, we do indeed arrive at null-compactified Minkowski space but in such a way as to keep the Kaluza-Klein tower fixed. As required, we arrive at $\text{Schr}(2n-2)$, with particle number $N$ identified by\footnote{We choose this sign for $N$, in line with the general NRCFT literature, since unitarity then requires $N\ge 0$, as discussed in Section \ref{subsec: unitarity}} $P_+ = -\frac{k}{R}N \to -R_+ N$, which does indeed have integer eigenvalues.

Indeed, this precise DLCQ limit of a $\frak{u}(1)\oplus \frak{su}(1,n)$ theory has been performed explicitly in the case $n=3$, both at the level of actions \cite{Mouland:2021urv} as well as correlators \cite{Lambert:2020zdc}.

\subsection{State-operator map} 

A deep and powerful result tool in the study of relativistic conformal field theory is the operator-state map, relating on one hand conformal primary operators, and on the other, eigenstates of the Hamiltonian of the theory on a sphere.
An analogous map exists in conventional non-relativistic  conformal field theories \cite{Nishida:2007pj}, which relates primary operators---defined in a way entirely analogous to the above---to eigenstates of the Hamiltonian augmented by a harmonic potential.

We will now show that construction applies in an almost identical way to the $\frak{u}(1)\oplus \frak{su}(1,n)$ theories of this work. Said another way, we verify that this operator-state map is not spoilt by the $\Omega$-deformation that parameterises our departure form the Schr\"odinger algebra of conventional non-relativistic CFT. Indeed, one may recover from our construction the familiar map of Nishida-Son in the Schr\"odinger limit as outlined in Section \ref{subsec: DLCQ}.

We approach the construction of our operator-state map from the perspective of automorphisms of the symmetry algebra, a well-established point of view in relativistic CFTs which has also recently been formulated for non-relativistic CFTs governed by the Schr\"odinger group \cite{Karananas:2021bqw}. 

Given some operator $\Phi(0)$ at the origin, we may define a state
\begin{align}
  \ket{\Phi} = \Phi(0) \!\ket{0}\ .
  \label{eq: op state}
\end{align}
Next, let us perform a Wick rotation in the symmetry algebra, defining $D=-i T$. Then, if $\Phi(0)$ has scaling dimension $\Delta$ under $T$, then
\begin{align}
  D\!\ket{\Phi} = D \Phi(0) \!\ket{0} = [D,\Phi(0)]\!\ket{0} = -i [T, \Phi(0)]\!\ket{0} = \Delta \Phi(0)\!\ket{0} = \Delta \!\ket{\Phi}\ ,
\end{align}
and thus $\ket{\Phi}$ has eigenvalue $\Delta$ under $D$. Then, just as with operators, we can use the ladder operators $\{H,K\}$ and $\{P_i, G_i\}$ to raise and lower the $D$ eigenvalue of $\ket{\Phi}$. For instance, $DH\!\ket{\Phi} = (\Delta+2)H\!\ket{\Phi}$, while $DG_i\!\ket{\Phi} = (\Delta-1)G_i\!\ket{\Phi}$.

We can consider $\ket{\Op}$ specifically for a primary operator $\Op$. This state then sits at the bottom of a semi-infinite tower of operators, since $K\!\ket{\Op}=0$ and $G_i \!\ket{\Op}=0$.

Thus, we have on one hand primary operators and their descendants, all with definite scaling dimension, and on the other hand, eigenstates of the operator $D=-iT$. Let us now however explore an alternative frame, related by a similarity transform on the Hilbert space and space of operators. As we shall see, this transformation, which can be seen as a non-relativstic analogue of the operator-state map of relativistic CFT, relates the spectra of $D$ with that of a combination of the form $\sim \left(H+K\right)$. In many physical examples, one can thus study the spectrum of $D$ by instead studying the dynamics of particles trapped in a confining potential provided by $K$ \cite{Nishida:2007pj,Nishida:2010tm}. We now show that this operator-state map present in Schr\"odinger invariant theories generalises to the theories studied here.

So let us consider transformed states and operators given by
\begin{align}\label{transOp}
  \ket{\bar{\Phi}} = e^{-\mu H}e^{\frac{1}{2\mu}K}\ket{\Phi},\qquad \bar{\Phi} = e^{-\mu H}e^{\frac{1}{2\mu}K} \Phi e^{-\frac{1}{2\mu}K} e^{\mu H}\ ,
\end{align}
for some constant $\mu$. Note, this transformation is clearly consistent with the identification (\ref{eq: op state}). In particular, for a \textit{primary} operator $\Op$ we have
\begin{align}
\label{eq:similarity transform}
  \ket{\bar{\Op}} = e^{-\mu H}e^{\frac{1}{2\mu}K} \Op(0)\!\ket{0}  = e^{-\mu H} \Op(0)\!\ket{0}\ ,
\end{align}
as is familiar from the usual non-relativistic operator-state map \cite{Nishida:2007pj}. Then, this defines an alternative map between on one hand the primary operators $\Op$ and their descendants, and on the other, towers of eigenstates of $\bar{D}$ generated by acting with the ladder pairs $\{\bar{H},\bar{K}\}$ and $\{\bar{P}_i,\bar{G}_i\}$.

Explicitly, the transformed operators under (\ref{transOp}) are
\begin{align}
\label{eq:barredgens}
  \bar{D} 		&=	\mu H+ \frac{1}{\mu}K			\nn\\
  \bar{H}		&=	\frac{1}{4\mu}\left(\mu H - \frac{1}{\mu}K + iT\right)				\nn\\
  \bar{K}		&=	- \mu 	\left(\mu H - \frac{1}{\mu}K - iT\right)				\nn\\
  \bar{P}_i		&=	\frac{1}{2}\left(P_i + \frac{1}{\mu}i G_i\right)				\nn\\
  \bar{G}_i		&= i\mu \left(P_i - \frac{1}{\mu}i G_i\right)	\ ,
\end{align}
while the remaining generators, the rotations and central charge, transform trivially as $\bar{B}=B$, $\bar{J}^\alpha = J^\alpha$ and $\bar{P}_+ = P_+$. A primary state $\ket{\bar{\Op}}$ then satisfies
\begin{align}
\label{eq:primary}
  \bar{D} \ket{\bar{\Op}} 		&= \Delta \ket{\bar{\Op}}				\nn\\
  \bar{P}_+ \ket{\bar{\Op}}		&= p_+ \ket{\bar{\Op}}					\nn\\
  \bar{B} \ket{\bar{\Op}}		&= - r_\Op[B] \ket{\bar{\Op}}			\nn\\
  \bar{J}^\alpha \ket{\bar{\Op}}	&= - r_\Op[J^\alpha] \ket{\bar{\Op}}		\nn\\
  \bar{G}_i \ket{\bar{\Op}}		&= 0									\nn\\
  \bar{K} \ket{\bar{\Op}}		&= 0 \ ,
\end{align}
while acting with $\bar{H}$ and $\bar{P}_i$ generates towers of descendents.

Up to normalisation these operators (\ref{eq:barredgens}) take the same form as in conventional non-relativistic CFT \cite{Nishida:2007pj,Doroud:2016mfv}, and thus automatically satisfy the same algebra in the $R\to\infty$ limit.

\subsection{Implications of unitarity}\label{subsec: unitarity}

If we assume unitarity in the original Minkowskian theory, then all states will have non-negative norm. Just as is the case of Lorentzian CFTs, we can use this assumption to place constraints on the eigenvalues of certain operators. The original Minkowskian symmetry generators were all Hermitian operators, but since we are interested in states quantised in the analogue of radial quantisation, we should instead consider the barred generators of \eqref{eq:barredgens}. Simple Hermiticity of the original generators implies for the barred generators the following reality conditions
\begin{align}
\begin{split}
\bar{D}^\dagger  &= \bar{D} \\ 
\bar{H}^\dagger  &= - \frac{1}{4\mu^2} \bar{K} \\
\bar{K}^\dagger  &= - 4 \mu^2 \bar{H} \\
\bar{P}_{i}^{\dagger} &= \frac{1}{2\mu} i \bar{G}_i \\
\bar{G}_{i}^{\dagger} &= 2 \mu i \bar{P}_i\ .
\end{split}
\end{align}
So let us now consider the primary state $\ket{\bar{\Op}}$ with data $\{\Delta,p_+, r_\Op[B],r_\Op[J^\alpha]\}$. Then, we have
\begin{align}
|\bar{H}\ket{\bar{\Op}} \!|^2 =  \bra{\bar{\Op}} \bar{H}^{\dagger} \bar{H} \ket{\bar{\Op}} =  - \frac{1}{4 \mu^2}\bra{\bar{\Op}} \!\big[ \bar{K}, \bar{H} \big]\! \ket{\bar{\Op}} = \frac{1}{4\mu^2} \bra{\bar{\Op}} \bar{D} \ket{\bar{\Op}} =  \frac{\Delta}{4 \mu^2}\braket{\bar{\Op}|\bar{\Op}}\ .
\end{align}
If the theory is unitary, all states have non-negative norm, implying that for primary states
\begin{align}
\Delta \geq 0\ .
\end{align}
Since $\bar{H}$ and $\bar{P}_i$ raise $\Delta$, this bound clearly holds for all descendants as well.
Let us similarly consider
\begin{align}
|\bar{P}_i|\bar{\Op} \rangle|^2 = \frac{i}{2 \mu} \bra{\bar{\Op}}\! \big[ \bar{G}_i, \bar{P}_i \big]\! \ket{\bar{\Op}} &=  \frac{1}{2 \mu} \bra{\bar{\Op}} \left( - \bar{P}_{+} + \frac{1}{2R} \frac{n+1}{n-1} \bar{B} + \beta^{\alpha}_{ii} \bar{J}^{\alpha} \right) \ket{\bar{\Op}}		\ ,
\end{align}
where $i$ is not summed over. But let us now sum over $i$, so as to exploit the tracelessness of $\beta^\alpha_{ij}$. Then, we arrive at
\begin{align}
\sum_i|\bar{P}_i|\bar{\Op} \rangle|^2 =  \frac{n-1}{ \mu}\left( - p_{+} - \frac{1}{2R} \frac{n+1}{n-1} r_\Op[{B}] \right) \braket{\bar{\Op}|\bar{\Op}}		\ ,
\end{align}
where recall that $r_\Op[B]\in \frac{1}{2}\mathbb{Z}$ is the charge of $\Op$ under a particular $U(1)$ rotation subgroup, while $p_+$ is its central charge, taking values in $\frac{1}{R}\mathbb{Z}$ in theories descended from $2n$-dimensional CFTs. Again imposing non-negative norms, we require
\begin{align}
\frac{M}{R}:= - p_+ - \frac{1}{2R} \frac{n+1}{n-1}r_\Op[B] \ge 0\ .
\end{align}
In particular, if $p_+\in \frac{1}{R}\mathbb{Z}$ then $M$ is rational. We can think of $M$ as playing a role analogous to particle number in conventional NRCFTs.

Note, we see that a scalar primary must have $p_+\le 0$. It is interesting to note that this condition appears to be manifestly realised in known supersymmetric interacting gauge theory examples of $\frak{su}(1,3)$ theories \cite{Lambert:2020zdc}, in a rather novel way. In particular, in such theories, $P_+$ is identified as instanton charge, and the dynamics are constrained such that only anti-instantons, corresponding to $p_+\le 0$, are allowed to propagate \cite{Lambert:2021mnu}.

We can use the positivity of $\Delta$ and $M$ to improve our bound for $\Delta$. In particular, for any primary with $M>0$, consider the norm \cite{Nishida:2010tm} 
\begin{align}
 \left|\left(\bar{H} -\frac{R}{2M} \sum_i \bar{P}_i \bar{P}_i\right)| \bar{\mathcal{O}} \rangle \right|^2 \geq 0\ .
\end{align}
This leads to the inequality
\begin{align}
(n-1)\Delta \left( \Delta - (n-1) \right) \geq 4(n-1) \left(  M^2  + \sum_{\alpha} r_{\mathcal{O}}[J^\alpha]^2 \right)\ ,
\end{align}
the right hand side is manifestly semi-positive, and we have already shown $\Delta$ is too, so one arrives at
\begin{align}
\Delta \geq n-1\ .
\end{align}
for any primary with $M>0$.
Since we have $2n-2$ spacial dimensions we see that, despite the $\Omega$ deformation, this bound agrees with the usual bound for theories with a Schr\"odinger symmetry algebra \cite{Nishida:2010tm}.

\section{Superconformal Extension in Six Dimensions}

We have thus far explored the reduction of symmetries of an even-dimensional conformal field theory when dimensionally reduced along a particular conformally-compactified direction. In six or fewer dimensions the conformal algebra admits extensions to various Lie superalgebras and thus it is natural to extend our analysis to determine the fate of supersymmetry under such dimensional reductions. In particular, any surviving supersymmetry constitutes a Lie superalgebra extension of $\frak{su}(1,n)$.

The dimensional reduction we have constructed is novel only for $n\ge 2$, while the starting $2n$-dimensional CFT can have supersymmetry only for $n=1,2,3$. 
 Motivated by a well-studied class of supersymmetric Lagrangian models with $\frak{su}(1,3)$ symmetry \cite{Lambert:2019fne,Lambert:2020jjm,Lambert:2020zdc,Lambert:2021mnu,Lambert:2021fsl}, we focus in this work on the case of $n=3$. It would be interesting however to explore the $n=2$ case in future work, where one would expect to recover an $\Omega$-deformed version of existing results on null reductions of four-dimensional superconformal algebras \cite{Sakaguchi:2008rx}.

\subsection{$\frak{osp}(8^*|4)\longrightarrow \frak{u}(1)\oplus \frak{osp}(6|4)$}

In six-dimensions the only choices for relativistic superconformal algebras are $D(4,1)$ and $D(4,2)$ corresponding to $\frak{osp}(8^*|2)$ and $\frak{osp}(8^*|4)$ respectively. In the following we cover the latter, the Bosonic part of which is $\frak{so}(6,2)\oplus \frak{so}(5)_R$, which we demonstrate is broken to $\frak{u}(1)\oplus \frak{osp}(6|4)$ by the conformal compactification. It is straightforward  to extend our discussion to $\frak{osp}(8^*|2)\to\frak{u}(1)\oplus \frak{osp}(6|2)$.

In Minkowski signature we choose conventions where all Bosonic generators are Hermitian, as before. Their commutation relations are the same as in Section \ref{subsec:reduc}. The R-symmetry generators have the standard form
\begin{align}
i[\hat{R}_{IJ}, \hat{R}_{KL}] = \delta_{J K} \hat{R}_{I L} + \delta_{I L} \hat{R}_{J K} - \delta_{I K} \hat{R}_{JL} - \delta_{JL} \hat{R}_{IK}\ .
\end{align}
with $I \in \{ 1,\dots , 5 \}$. The Fermionic generators are six-dimensional symplectic-Majorana-Weyl Fermions. The reality condition as applied above is 
\begin{align}
\hat{Q}_{\alpha A} = i \Omega_{AB} (\Gamma_{0})_{\alpha}^{\ \ \beta} C_{\beta \gamma} (\hat{Q}_{\gamma B})^{\dagger}\ ,
\end{align}
and similar for $\hat{S}$, again this is in Minkowski signature, and the dagger here is not transposing spin indices (spinors see it as just complex conjugation). $\Omega^{AB}$ and $C^{\alpha \beta}$ are the five  and six-dimensional charge conjugation matrices, with $A \in \{ 1,\dots , 4\}$ and $ \alpha \in \{ 1 , \dots , 8 \}$.\footnote{ Note that $\Omega_{AB}$ should not be confused with $\Omega_{ij}$ which we used in the coordinate transformation (\ref{eq: coordinate transformation}). To ameliorate  this problem we will always explicitly write the indices. } The $\hat{Q}$ and $\hat{S}$ have opposite chirality under $\Gamma_* = \Gamma_{012345}$.
\begin{align}
\Gamma_* \hat{Q} = -\hat{Q}, \quad \Gamma_* \hat{S} = \hat{S}\ .
\end{align}
Again we wish to find the maximal subalgebra of all elements that commute with the element $P_{+}$, defined in terms of the six-dimensional (hatted) operators as 
\begin{align}
P_{+} = \hat{P}_{+} + \frac{1}{4} \Omega_{ij} \hat{M}_{ij} + \frac{1}{8 R^2}\hat{K}_{-}\ .
\end{align}
We find that $3/4$ of the supercharges commute with $P_+$. Precisely which set of supercharges this is depends on whether $\Omega_{ij}$ is self-dual or anti-self-dual; without loss of generality, let us choose the latter case. Then, letting a $\pm$ subscript denote chirality under $\Gamma_{05}$, the commuting supercharges are
\begin{align}
\begin{split}
Q_{-A} &:= \Gamma_{-} \hat{Q}_{-A} \\
S_{+A} &:= \Gamma_{+} \hat{S}_{+A} \\
\Theta_{-A} &:= \frac{1}{4} \Big( R \Omega_{ij} \Gamma_{ij} \hat{Q}_A + \frac{1}{R} \Gamma_{-} \hat{S}_A \Big)\ .
\end{split}
\label{eq:newgens}
\end{align}
The alternative case, where $\Omega_{ij}$ is self-dual, is found simply by swapping all $\Gamma_{05}$ chiralities. Then, their commutation relations with the bosonic generators are
\begin{align}
\begin{split}
i[Q_{-}, H]		&=	0 \\
i[Q_{-}, P_i]		&=	0 \\
i[Q_{-}, B]			&=	0 \\
i[Q_{-}, J^\alpha]		&=	\frac{1}{4} L_{ij}^\alpha \Gamma_{ij} Q_{-} \\
i[Q_{-}, T]		&=	Q_{-} \\
i[Q_{-},G_{i}]	&=	 -R \Omega_{ij}\Gamma_{j} \Theta_{-} \\
i[Q_{-}, K]		&= 	-\frac{1}{2} \Gamma_- S_{+} \\
i[Q_{-A}, R_{IJ}]	&=	\frac{1}{2}(\tilde{\Gamma}_{IJ})_{A}^{\ \ B} Q_{-B}
\end{split}
\begin{split}
i[S_{+}, H]		&=	- \Gamma_+ Q_{-} \\
i[S_{+}, P_i]		&=	 R \Omega_{ij} \Gamma_{+j} \Theta_{-} \\
i[S_{+}, B]			&=	0 \\
i[S_{+}, J^\alpha]		&=	\frac{1}{4}L_{ij}^\alpha \Gamma_{ij} S_{+} \\
i[S_{+}, T]			&=	-S_{+} \\
i[S_{+}, G_{i}]	&=	0 \\
i[S_{+}, K]		&= 	0 \\
i[S_{+A}, R_{IJ}]	&=	\frac{1}{2}(\tilde{\Gamma}_{IJ})_{A}^{\ \ B} S_{+ B}
\end{split}
\begin{split}
\ i[\Theta_{-}, H]		&=	0 \\
\ i[\Theta_{-}, P_i]		&=	\frac{1}{2R} \Gamma_i Q_{-} \\
\ i[\Theta_{-}, B]			&=	\frac{1}{4}R \Omega_{ij} \Gamma_{ij} \Theta_{-} \\
\ i[\Theta_{-}, L^\alpha]		&=	0 \\
\ i[\Theta_{-}, T]			&=	0 \\
\ i[\Theta_{-}, G_{i}]		&=	\frac{1}{4R}\Gamma_{- i} S_{+} \\
\ i[\Theta_{-}, K]		&= 	0 \\
\ i[\Theta_{- A}, R_{IJ}]	&=	\frac{1}{2}(\tilde{\Gamma}_{IJ})_{A}^{\ \ B} \Theta_{-B}\ ,
\end{split}
\end{align}
while we have anti-commutators
\begin{align}
i\{ Q_{- \alpha A}, Q_{- \beta B} \} &= - 4 ( \Gamma_- \Pi_- C^{-1})_{\alpha \beta} \Omega_{AB} H \nn\\ 
i\{ S_{+ \alpha A},
S_{+ \beta B} \} &= - 8 ( \Gamma_+ \Pi_+ C^{-1})_{\alpha \beta} \Omega_{AB} K \nn\\
i\{ \Theta_{- \alpha A}, \Theta_{- \beta B} \} &= -2 (\Gamma_{-} \Pi_{+} C^{-1})_{\alpha \beta} \Omega_{AB} (P_+ - B/R) - 1/4 \Omega_{i j}  (\Gamma_{-} \Gamma_{ij} \Pi_{+} C^{-1})_{\alpha \beta} (\tilde{\Gamma}_{IJ}\Omega^{-1})_{AB} R_{IJ} \nn\\
i\{ Q_{- \alpha A}, S_{+ \beta B} \} &= -2(\Gamma_{-} \Gamma_{+} \Pi_{+} C^{-1})_{\alpha \beta} \big ( \Omega_{AB} T + (\tilde{\Gamma}_{IJ}\Omega^{-1})_{AB} R^{IJ} \big ) - 1/2 ( \Gamma_{ij} \Gamma_{-} \Gamma_{+} \Pi_{+} C^{-1})_{\alpha \beta} \Omega_{AB} L^{\alpha}_{ij} J^\alpha \nn\\
i\{ Q_{- \alpha A}, \Theta_{- \beta B} \} &= -2 R \Omega_{ij} ( \Gamma_{-} \Gamma_{j}  \Pi_{+} C^{-1})_{\alpha \beta} \Omega_{AB} P_i \nn\\
i \{ S_{+ \alpha A}, \Theta_{- \beta B} \} &= -2 R \Omega_{ij} ( \Gamma_{+} \Gamma_{-} \Gamma_{j} \Pi_{+} C^{-1})_{\alpha \beta} \Omega_{AB} G_{i}\ ,
\end{align}
where we have defined the projectors $\Pi_{\pm} = 1/2 (1 \pm \Gamma_{*})$. 

Thus there are $50 = 1+15+10+24$ Bosonic generators corresponding to the central extension, $\frak{su}(1,3)$ and $\frak{so}(5)$, as well as $3\times 8=24$ Fermionic generators. The superalgebra is a realisation of $\frak{u}(1)\oplus \frak{osp}(6|4)$.

The Fermionic generators can also be transformed by \eqref{eq:similarity transform}, which yields
\begin{align}
\begin{split}
\bar{Q}_{-A} &= - \frac{1}{2}i Q_{-A} - \frac{1}{4\mu} \Gamma_{-} S_{+A} \\
\bar{S}_{+A} &= - i S_{+A} + \mu \Gamma_{+} Q_{-A} \\ 
\bar{\Theta}_{-A} &= - i\Theta_{-A}\ .
\end{split}
\end{align}
Taking  a symplectic-Majorana-Weyl reality condition for the six-dimensional spinors we find the following Hermiticity properties for the barred generators
\begin{align}
\label{eq:femireality}
\begin{split}
\bar{Q}_{-A}^{\dagger} &= \frac{1}{4\mu} \Omega_{AB} C \Gamma_{0} \Gamma_{-} \, \bar{S}_{+B} \\ 
\bar{S}_{+A}^{\dagger} &= -2 \mu \, \Omega_{AB} C \Gamma_{0} \Gamma_{+} \, \bar{Q}_{-B} \\ 
\bar{\Theta}_{-A}^{\dagger} &= \Omega_{AB} C \Gamma_{0} \, \bar{\Theta}_{-B}
\ .\end{split}
\end{align}

Rather unusually for such algebras, along with a pair of of Fermionic generators that raise and lower the eigenvalue of $T$, namely the $Q_{-}$ and $S_{+}$, we also have  generators that do not change this eigenvalue; $\Theta_{-}$. We can see that while $Q_{-}^2 \sim H$ and $S_{+}^2 \sim K$, $\Theta_{-}^2 \sim M + (\text{R-sym})$. This has an interesting effect on the usual process of defining super-conformal primaries. Let a superconformal primary be any state satisfying  \eqref{eq:primary} and that is further annihilated by $ \bar{S}_{+} $. Then, given such a state with scaling dimension $\Delta$, one can form a family of primary states all of dimension $\Delta$, by acting repeatedly with different $\bar{\Theta}_{-}$ operators. These states can be seen to be primary as acting again with $\bar{K},\bar{G}_{i}$ or $\bar{S}_{+}$ still annihilates the state
\begin{align}
\begin{split}
& \bar{K} \bar{\Theta}_{- \alpha A} | \bar{\mathcal{O}} \rangle = \left[\bar{K}, \Theta_{- \alpha A}]| \bar{\mathcal{O}} \right] \rangle = 0 \\
& \bar{G}_{i} \bar{\Theta}_{- \alpha A} | \bar{\mathcal{O}} \rangle = \left[\bar{G}_{i}, \Theta_{- \alpha A}]| \bar{\mathcal{O}} \right] \rangle \sim  \bar{S}_{+} | \bar{\mathcal{O}} \rangle = 0 \\
&  \bar{S}_{+ \beta B} \bar{\Theta}_{- \alpha A} | \bar{\mathcal{O}} \rangle = \{ \bar{S}_{+ \beta B}, \bar{\Theta}_{- \alpha A} \} \sim \bar{G}_{i}  | \bar{\mathcal{O}} \rangle = 0\ .
\end{split}
\end{align}
It follows inductively that any number of $\bar
{\Theta}_{-}$ times a primary is still primary. One cannot form infinitely many of these states as each $\Theta_-$ is nilpotent, so each super conformal primary belongs to a family of such states, an original bosonic state, plus those that follow from the  action of $\bar{\Theta}_{-}$.
Unitarity bounds for superconformal primary states can also be calculated using \eqref{eq:femireality}. We consider first the norm
\begin{align}
|\bar{\Theta}_{- \alpha A} | \bar{\mathcal{O}} \rangle |^2 \geq 0\ ,
\end{align}
which leads to the inequality 
\begin{align}
\langle \bar{\mathcal{O}} | {\bar{\Theta}^{\dagger}}_{\! \! - \alpha A } \bar{\Theta}_{- \alpha A} | \bar{\mathcal{O}} \rangle = \sum_{\beta, B} \Omega^{A B} (C\Gamma_{0})^{\alpha \beta}  \langle \bar{\mathcal{O}} | \bar{\Theta}_{- \beta B} \bar{\Theta}_{- \alpha A}  | \bar{\mathcal{O}} \rangle \geq 0\ .
\end{align}
Summing again on $\alpha$ and $A$ symmetrises on simultaneous exchange of $\alpha, \beta$ and $A,B$, allowing us to replace the product with the anticommutator. This then simply reproduces the earlier bound $M \geq 0$. 

A more interesting bound is found from the norm
\begin{align}
|Q_{- \alpha A} | \bar{\mathcal{O}} \rangle |^2 \geq 0
\end{align}
which leads to
\begin{align}
\begin{split}
\langle \bar{\mathcal{O}} | {\bar{Q}^{\dagger}}_{\! \! - \alpha A } \bar{Q}_{- \alpha A} | \bar{\mathcal{O}} \rangle = \frac{1}{4 \mu}\sum_{\beta B} \Omega^{A B} (C \Gamma_{0} \Gamma_{-})^{\alpha \beta} \langle \bar{\mathcal{O}} |  \{ {\bar{S}}_{+ \beta B } \bar{Q}_{- \alpha A} \} | \bar{\mathcal{O}} \rangle \geq 0 \ ,
\end{split}\end{align}
and implies 
\begin{align}
\begin{split}\frac{1}{4 \mu}\sum_{\beta B} \Omega^{A B} (C \Gamma_{0} \Gamma_{-})^{\alpha \beta} \langle \bar{\mathcal{O}} |-2(\Gamma_{-} \Gamma_{+} \Pi_{+} C^{-1})_{\alpha \beta} \big ( \Omega_{AB} \bar{T} + (\tilde{\Gamma}_{IJ}\Omega^{-1})_{AB} \bar{R}^{IJ} \big ) \\ - 1/2 ( \Gamma_{ij} \Gamma_{-} \Gamma_{+} \Pi_{+} C^{-1})_{\alpha \beta} \Omega_{AB} L^{\alpha}_{ij} \bar{J}^\alpha  | \bar{\mathcal{O}} \rangle \geq 0\ .
\end{split}
\end{align}
Since $\bar{S}_{+}$ annihilates primaries, we do not need to symmetrise to replace the product with the anti-commutator. For example when we pick $\alpha = 5$ and $A=1$ we find
\begin{align}
\Delta \geq r_{\mathcal{O}}[J^1] + r_{\mathcal{O}}[R_{12}]+r_{\mathcal{O}}[R_{34}]\ ,
\end{align}
where we defined $ \bar{R}_{IJ} | \bar{\mathcal{O}} \rangle = -r_{\mathcal{O}}[R_{IJ}] |\bar{\mathcal{O}}\rangle $.

It is interesting to note that, up to a choice of real form for the respective algebras, the reduction of symmetry from the six-dimensional $(2,0)$ superalgebra down to to centraliser of $P_+$ is identical to the symmetry breaking pattern of the classical ABJM theory, which realises manifestly only a particular subalgebra of the full three-dimensional $\mathcal{N}=8$ superconformal algebras. A detailed discussion of this correspondence, including its holographic origin, can be found in \cite{Mouland:2021urv}.

\section{Free Fields in Various Dimensions}

In this section we want to discuss examples of  field theories in $(2n-1)$-dimensions with $SU(1,n)$ symmetry. Our examples will be obtained by the conformal compactification of a $2n$-dimensional free conformal theory.   We will include the entire Kaluza-Klein tower in our discussion but as the $SU(1,n)$ symmetry acts on each level independently one is also free to truncate the actions to only include fields of particular levels. Interacting versions of these theories can also be constructed by starting with an interacting conformal field theory, for example by considering the reduction of non-Abelian theories.   In the interests of clarity we will not consider these here. 
 
\subsection{Scalars in $2n-1$ dimesions}

To begin we consider a free real scalar in $(1+1)$-dimensions, {\it i.e.} $n=1$. As we will see this case is special, yet familiar. In particular we start with the action for a real scalar field:
\begin{align}
S_{2D} = \frac{1}{ g^2}\int d\hat x^+ d\hat x^-   \hat \partial_+{\hat \phi} \hat \partial_- \hat \phi    \ ,
\end{align}
where in this simple case 
\begin{align}
\hat x^+ &= 2R \tan (x^+/2R) \, , \nonumber\\ 
\hat x^- &= x^- \ .
\end{align}Since $\hat \phi$ has scaling dimension zero we simply find $\hat \phi=\phi$. Thus we expand
\begin{align}
\hat\phi=	     \sum_{k\in\mathbb Z} e^{ikx^+/R}\phi^{(k)}(x^-)\ .
\end{align}
Note that    $\phi^{(-k)}={\phi^{(k)}}^\dag$.  In a more standard treatment of the two-dimensional scalar one would solve the equations of motion which sets  $\phi^{(k)}$, for $k\ne 0$,  to constant left-moving oscillators whereas $\phi^{(0)}(x^-)$ is expanded in terms of right moving oscillators.  One might also consider including winding modes but we will not do so  here as the spatial direction is not compact.

Substituting into the action we find
\begin{align}
S_{2D} \label{2Dscalar}
&= \frac{2\pi   }{g^2} \sum_{k\in\mathbb Z}\int dx^-\ ik\partial_-\phi^{(k)} \phi^{(-k)} \nonumber\\
&=  \frac{2\pi  ik }{g^2} \sum_{k> 0}\int dx^-\left( \partial_-\phi^{(k)} {\phi^{(k)}}^\dag- \phi^{(k)} \partial_-{\phi^{(k)}}^\dag \right)\ .
\end{align}

 By construction the $SU(1,1)$ symmetry separately on each of the fields $\phi^{(k)}$   at fixed $k\in\mathbb Z$. Translations act as 
  \begin{align}
 \phi^{(k)} \to \phi^{(k)} + \epsilon\partial_-\phi^{(k)}\ .	
 \end{align}
 The Liftshitz scaling $T$ is simply 
\begin{align}
\phi^{(k)} \to \phi^{(k)} + \lambda x^-\partial_-\phi^{(k)}	\ .
\end{align} 
Finally the special conformal transformation $K_+$ acts as:
\begin{align}
\phi^{(k)}\to \phi^{(k)} + 2\kappa ({x^-})^2\partial_-\phi^{(k)}\ .
\end{align} 
One can readily check that these are indeed symmetries to first order. 

However we see that they can be extended  to
\begin{align}
\phi^{(k)}\to \phi^{(k)} + f (x^-)\partial_-\phi^{(k)}	\ ,
\end{align} 
for any function $f(x^-)$. Taking $\kappa$ constant, linear and quadratic leads to the $H$, $T$ and $K$ generators, respectively. In fact this is simply the action of one-dimensional diffeomorphisms and therefore yields an infinite-dimensional symmetry group with generators
\begin{align}
	L_n = (x^-)^{n+1}\partial_-\ .
\end{align}
These satisfy the Witt algebra
\begin{align}
[L_m,L_n] = (n-m)L_{m+n}	\ ,
\end{align}
where $H=L_{-1},T=L_0,K=L_1$ form a finite dimensional subalgebra.  
However just as in the familiar case of the string worldsheet in the quantum theory, where we must normal order the operators $\phi^{(k)}$, we will generate a central charge $c=1$.

Let us now consider a free real scalar obtained from reduction from $D=2n$:\footnote{There is also a coupling to the spacetime Ricci scalar but since we are working on a conformally flat metric, this term vanishes.}
\begin{align}
S 
& = -\frac{1}{2g^2} \int d^{2n-2}x dx^+dx^-\sqrt{-\det (\hat g)}
\hat g^{\mu\nu}\partial_\mu \hat\phi  \partial_\nu \hat\phi\ ,
\end{align}
where $\hat g_{\mu\nu}$ is the metric in (\ref{5dmink}). As before we perform a conformal rescaling to the metric  (\ref{eq: conformally compactified metric}) to obtain
\begin{align}
S = &  \frac{1}{2g^2} \int d^{2n-2}x dx^+dx^-\left[  2 \partial_+\phi \partial_-\phi  -\frac{|\vec x|^2}{4R^2}\partial_-\phi\partial_-\phi + \Omega_{ij}x^j\partial_-\phi \partial_i\phi-\partial_i  \phi \partial_i  \phi\right]\ .
\end{align}
Next  we expand
\begin{align}
	\hat\phi &= (\cos(x^+/2R))^{n-1}\phi\nonumber\\
& = 	(\cos(x^+/2R))^{n-1}\sum_k e^{ikx^+/R}\phi^{(k)}(x^-,x^i)\ .	
\end{align} 
Note that we do not necessarily require that $k\in\mathbb Z$. In fact if we impose that $\hat\phi$ is periodic on $x^+\in[-\pi R,\pi R]$ then we require $k$ to be integer for $n$ odd but half integer for $n$ is even. In this way we find
\begin{align}\label{DDscalar}
S =  \frac{\pi  R }{g^2} \sum_{k} \int d^{2n-2}x  dx^-\Big[ \frac{2ik}{R} \phi^{(k)}\partial_-\phi^{(-k)}&-\frac{|\vec x|^2}{4R^2 }\partial_-\phi^{(k)}\partial_-\phi^{( -k)}  \nonumber\\
 & + \Omega_{ij}x^j\partial_i\phi^{(k)}\partial_-\phi^{(-k)} -\partial_i  \phi^{(k)}  \partial_i  \phi^{(-k)}\Big]	\ .
\end{align}
As discussed this action admits  an $SU(1,n)$ spacetime symmetry acting on each level $k$ independently.

\subsection{Fermions in $2n-1$ dimensions}

Let us consider the reduction of a Fermion. Starting in $2n$ dimensions we have
\begin{align}
S & = \frac{i}{2g^2}\int d^{2n-2}xdx^+dx^-{\det(\hat e)} \bar{\hat \psi }\hat e^{\mu}{}_{\underline \nu} \gamma^{\underline \nu}  {\hat \nabla}_\mu \hat\psi\ .
\end{align}
Here  $\hat e_\mu{}^{\underline \nu}$ is the vielbein of the metric (\ref{5dmink}) and $\gamma^{\underline \nu}$ the $n$-dimensional $\gamma$-matrices of the tangent space. To keep our discussion general we do not impose any conditions on $\hat\psi$ and treat it as a Dirac spinor. In particular we assume that $\bar{\hat \psi} = \hat\psi^\dag  \gamma_{\underline 0}$.

We see that $\hat \psi$ has conformal dimension $n-1/2$. Thus we expand
\begin{align}
\hat\psi(x^+,x^-,x^i) = \cos^{n-1/2}(x^+/2R)\sum_k e^{ikx^+/R}\psi^{(k)}(x^-,x^i) 	\ .
\end{align}
Note that we do not necessarily impose ${\psi^{(k)}}^\dag = \psi^{(-k)}$.  

This leads to the reduced action
\begin{align}
S  = \frac{\pi R}{g^2} \sum_k\int d^{2n-2}x dx^- \Big(&-	i\bar\psi^{(k)} \gamma_+\partial_-\psi^{(-k)}+ i\bar\psi^{(k)} \gamma_i\partial_i\psi^{(-k)}+\frac{i}{2}\Omega_{ij}x^j\bar\psi^{(k)}\gamma_i\partial_-\psi^{(-k)}\nonumber\\
&+\frac{k}{R}\bar\psi^{(k)}\gamma_-\psi^{(-k)}+\frac{i}{8}\Omega_{ij}\bar\psi^{(k)}\gamma_{ij}\gamma_-\psi^{(-k)}\Big)\ ,
\end{align}
where now $\gamma_-,\gamma_+,\gamma_i$ are simply the $\gamma$-matrices of flat spacetime ({\it i.e.} the same as $\gamma_{\underline \nu}$).

We it is helpful to split $\psi^{(-k)} = \psi^{(-k)}_++\psi^{(-k)}_-$ where $\gamma_{-+}\psi^{(-k)}_{\pm}=\pm\psi^{(-k)}_{\pm}$. To clean things up we let
\begin{align}
\lambda^{(k)} = \tfrac12(1+\gamma_{-+})\psi^{(k)}\qquad \chi^{(k)}	=\tfrac12(1-\gamma_{-+})\psi^{(k)}\ .
\end{align}
The action is then
\begin{align}
S  =  &\frac{\sqrt{2}\pi R}{g^2} \sum_k\int d^{2n-2}x dx^- \Big(-	i{\lambda^{(k)}}^\dag\partial_-\lambda^{(-k)} + i{\lambda^{(k)}}^\dag \gamma_0\gamma_i\partial_i\chi^{(-k)} +  i{\chi^{(k)}}^\dag\gamma_0\gamma_i\partial_i\lambda^{(-k)} \nonumber\\
& +\frac{k}{R}{\chi^{(k)}}^\dag\chi^{(-k)}+\frac{i}{2}\Omega_{ij}x^j{\lambda^{(k)}}^\dag \gamma_0\gamma_i\partial_-\chi^{(-k)}+\frac{i}{2}\Omega_{ij}x^j{\chi^{(k)}}^\dag\gamma_0\gamma_i\partial_-\lambda^{(-k)} +\frac{i}{8}\Omega_{ij}{\chi^{(k)}}^\dag\gamma_{ij}\chi^{(-k)} \Big)\ .
\end{align}
Note that the last term essentially leads to a shift in $k$ for some components of $\chi^{(k)}$, depending on the eigenvalue of $i\gamma_{ij}\Omega_{ij}$. It can also vanish if $\hat\psi$ satisfies an additional chirality constraint. In addition we have not specified the range of $k$. Indeed $\hat\psi$  contains both Weyl chiralities and in principle we could  take different choices of $k$ for the two chiralities. This is analogous to the various spin structures of the NS-R string. 

Finally we observe that in one-dimension we  simply find 
\begin{align}
S = \frac{\sqrt{2}\pi R}{g^2} \sum_k\int dx^- \Big(-	i{\lambda^{(k)}}^\dag  \partial_-\lambda^{(k)}	+\frac{k}{R}\chi^{(k)}\chi^{(k)}\Big)\ .
\end{align}
One again the action has an infinite dimensional symmetry generated by $L_{n}$ provided that the $\chi^{(k)}$ are invariant. Furthermore we will encounter a central charge $c=1/2$ once we normal order the fields in the quantum theory.

\subsection{A 1-form gauge field in $3$-dimensions}

Let us start with a  free four-dimensional Maxwell gauge field 
\begin{align}
S &= -\frac{1}{4e^2}\int d^4\hat x\sqrt{-\hat g}\hat g^{\mu\lambda}\hat g^{ \nu\rho} \hat F_{\mu\nu}
\hat F_{\lambda\rho}\ ,
\end{align}
where $\hat g_{\mu\nu}=\eta_{\mu\nu}$ is flat Minkowski space.
Our first step is to change coordinates, conformally rescale the metric to $g_{\mu\nu}$ and  Fourier expand
\begin{align}
\hat A_{ \mu} 
&=\cos( x^+/2R)A_{\mu}\nonumber\\
&= \cos( x^+/2R)\sum_{k\in {\mathbb Z}+\tfrac12} e^{ik x^+/R}A^{(k)}_{ \mu}	(x^-,x^i)\ .
\end{align}
Performing the integral over $x^+$ we obtain
\begin{align}
S = \frac{2\pi R}{ e^2}\sum_k\int dx^-d^2x	&\Big[\frac12\left(\frac{ik}{R}A^{(k)}_--\partial_- A^{(k)}_+\right)\left(-\frac{ik}{R}A^{(-k)}_--\partial_-A^{(-k)}_+\right)-\frac14 {\cal F}^{(k)}_{ij}{\cal F}^{(-k)}_{ij}\nonumber\\
&+\left(\frac{ik}{R}{\cal A}^{(k)}_i-\partial_iA^{(k)}_+  +\frac12\Omega_{il}x^l\partial_- A^{(k)}_+\right) F^{(-k)}_{-i} 
\Big]\ ,
\end{align}
where
\begin{align}
{\cal A}^{(k)}_i &= A^{(k)}_i-\frac12 \Omega_{ij}x^jA_-^{(k)}\nonumber\\
{\cal F}^{(k)}_{ij} &= F^{(k)}_{ij} - \frac12\Omega_{im}x^mF^{(k)}_{-j}	 +\frac12\Omega_{jm}x^mF^{(k)}_{-i}	\ ,
\end{align}
and we must identify $A^{(-k)}_{  \mu} = (A^{(k)}_{ \mu})^\dag$.
One could also consider a non-Abelian gauge field but we will not do this here.

\subsection{A 2-form gauge field in $5$-dimensions}

Finally we consider a free tensor in six-dimensions:
\begin{align}
S &= -\frac{1}{2\cdot 3!g^2}\int d^6\hat x\sqrt{-\hat g}\hat g^{\mu\rho}\hat g^{\nu\sigma}\hat g^{ \lambda\tau} \hat H_{\mu\nu\lambda}
\hat H_{\hat\rho\sigma\tau}\ ,
\end{align}
where $\hat H_{\mu\nu\lambda}= 3\partial_{[\mu}\hat B_{\nu\lambda]}$.
We then conformally rescale the metric to $g_{\mu\nu}$ and  Fourier expand
\begin{align}
\hat B_{ \mu\nu} 
&=\cos^2( x^+/2R)B_{\mu\nu}\nonumber\\
&= \cos^2( x^+/2R)\sum_{k\in {\mathbb Z}} e^{ik x^+/R}B^{(k)}_{ \mu\nu}	(x^-,x^i)\ ,
\end{align}
with $(B^{(k)}_{ \mu\nu}	)^\dag = B^{(-k)}_{ \mu\nu}	$
and reduce to $4+1$ dimensions. In particular if we let $C^{(k)}$ with components
\begin{align}
C^{(k)}_-=B^{(k)}_{+-}\qquad C^{(k)}_i=B^{(k)}_{+i}\ ,
\end{align}
be a five-dimensional one-form with 2-form field-strength $(G^{(k)}_{-i},G^{(k)}_{ij})$
then we find 
\begin{align}
S = -\frac{2\pi R}{ g^2}\sum_k\int d^4 xdx^- & \left( \frac12\left(  \frac{ik}{R}B^{(k)}_{-i}-G^{(k)}_{-i}\right)\left(  \frac{ik}{R}B^{(-k)}_{-i}+G^{(-k)}_{-i}+\frac12\Omega_{kl}x^l H^{(-k)}_{-ki}\right)\right. \nonumber\\
&\left. +\frac{1}{2}\left({\cal G}^{(k)}_{ij}-\frac{ik}{R}{\cal B}^{(k)}_{ij}\right)H^{(-k)}_{-ij}  +\frac{1}{2\cdot 3!}{\cal H}^{(k)}_{ijk}{\cal H}^{(-k)}_{ijk}\right)\ .
\end{align}
Here $(H^{(k)}_{-ij}, H^{(k)}_{ijk})$ are the 3-form field-strength components of the five-dimensional 2-form $(B^{(k)}_{-i},B^{(k)}_{ij})$ and 
\begin{align}
{\cal B}_{ij} &= B_{ij}-\frac14 \Omega_{il}x^lB_{-j}	-\frac14 \Omega_{jl}x^lB_{i-}	\nonumber\\
{\cal G}_{ij} &= G_{ij}-\frac14 \Omega_{il}x^lG_{-j}	-\frac14 \Omega_{jl}x^lG_{i-}	\nonumber\\
{\cal H}_{ijk} & = H_{ijk} -\frac12 \Omega_{il}x^lH_{-jk}-\frac12 \Omega_{jl}x^lH_{i-j}-\frac12 \Omega_{kl}x^lH_{ij-}\ .
\end{align}

 \section{Recovering $2n$-Dimensional Physics}
 
 In this section we would like to see how, by considering the entire Kaluza-Klein tower, we can reconstruct the correlation functions of the $2n$-dimensional theory that we started with. Since there are additional complications that enter when the field has a non-trivial Lorentz transformation we  will restrict our attention here to scalar fields. 
 
 \subsection{From one to two dimensions}
 
 Let us start with a tower of scalar fields in one-dimension that are obtained from a two-dimensional scalar as given in (\ref{2Dscalar}).
 We can read off from the action (\ref{2Dscalar}) that the correlation functions are of the form ($k>0$)  
\begin{align}
\langle 0|\phi^{(k)}(x_2^-)\phi^{(l)}(x_1^-)|0\rangle &=   \frac{g^2}{ 4\pi k}{\Theta(x_2^--x_1^-)} \delta_{k,-l}	\ . 	
\end{align}

Let us try to compute a two-point function of the original two-dimensional theory.  If we try to compute $\langle0|\phi(\hat x_2)\phi (\hat x_1)|0\rangle$ we do not find a translationally invariant answer as $\phi$ is not a conformal primary. Thus instead we consider the  correlator
  \begin{align}
	\langle 0|\partial_+ \phi (x_2)\partial_+\phi (x_1)|0\rangle &= -
	\sum_{k  }\sum_{l   } e^{ikx^+_2/R} e^{ ikx^+_1/R}\frac{kl}{R^2}\langle 0|  \phi^{(k)}(x_2^-) \phi^{(l)}(x_1^-)|0\rangle\nonumber\\
	&=  \frac{g^2}{4\pi  R^2}\sum_{k}ke^{ik( x_2^+-x_1^+)/R} \Theta(x_2^--x_1^-)\ .
\end{align}
We note that the sum over the Fourier modes is ill-defined. We can consider an $i\varepsilon$ prescription $x_2^+-x^+_1\to x^+_2-x^+_1+i\varepsilon$ but this will only work for the $k>0$ contributions (or similarly $x^+_2-x^+_1\to x^+_2-x^+_1-i\varepsilon$ will only work for $k<0$). To obtain a finite answer we therefore impose the additional condition
\begin{align}
 \phi^{(k)}(x^-)|0\rangle =0\qquad k>0 \ .
\end{align} 
This condition is of course familiar from the usual Hamiltonian treatment where $\phi^{(k)}$ are the left moving oscillators.
Thus we are left with 
\begin{align}
\langle 0|\partial_+ \phi (x_2)\partial_+\phi (x_1)|0\rangle 	= \frac{g^2}{2\pi  R^2}\sum_{k=0}^\infty k e^{ik(x^+_2-x^+_1+i\varepsilon)/R}  \Theta(x_2^--x_1^-)\ .
\end{align}
To evaluate this we note that
\begin{align}
	\sum_{k=0}^\infty e^{ik(x+i\varepsilon)/R}  = \frac{1}{1-e^{i (x+i\varepsilon)/R}}\ ,
\end{align}
and differentiating gives
\begin{align}
		\sum_{k=0}^\infty ke^{ik(x+i\varepsilon)/R}  &= \frac{e^{i (x+i\varepsilon)/R}}{(1-e^{i (x+i\varepsilon)/R})^2}\nonumber\\
		& = -\frac{1}{4\sin^2((x+i\varepsilon)/2R)} \ .
\end{align}
 Continuing we find (setting $\varepsilon=0$)
\begin{align}
	\langle 0| \partial_+ \phi(x_2)\partial_+\phi (x_1)|0\rangle  
	&=  -\frac{g^2}{4\pi  }\frac{1}{4R^2}\left[\frac{1}{\sin^2((x^+_2-x^+_1)/2R)}\right]\Theta(x_2^--x_1^-) \ .
\end{align}
On the other hand we have
\begin{align}
	\hat x^+_2-\hat x^+_1 &= 2R\tan(x^+_2/2R)-2R\tan(x^+_1/2R)\nonumber\\
	&={2R\tan((x^+_2-x^+_1)/2R)}\left[{1+\tan(x^+_2/2R)\tan(x^+_1/2R)}\right]\nonumber\\
	& =  \frac{2R\sin((x^+_2-x^+_1)/2R)}{\cos(x^+_2/2R)\cos(x^+_1/2R)} \ ,
\end{align}
and hence
\begin{align}
	\langle 0|\partial_+ \phi(x_2)\partial_+\phi (x_1)|0\rangle  
	&=  -\frac{g^2}{4\pi  } \left[\frac{1}{\cos^2(x^+_2/2R)\cos^2(x^+_1/2R)}\frac{1}{(\hat x^+_2-\hat x^+_1 )^2}\right]\Theta(x_2^--x_1^-) \ ,
\end{align}
which in terms of the original coordinates is
\begin{align}
	\langle 0|\hat \partial_+ \hat \phi (\hat x_2)\hat \partial_+ \hat \phi (\hat x_1)|0\rangle  
	&= -\frac{g^2}{4\pi  }  \frac{1}{(\hat x^+_2-\hat x^+_1 )^2} \Theta(\hat x_2^--\hat x_1^-) \ ,
\end{align}
which is the correct propagator for the two-dimensional theory.

 It is clear that from this treatment  we will never be able to reconstruct the right-moving sector as only $\Theta(x_2^--x^-_1)$ appears. This is in part due to our choice of quantisation. By choosing $x^-$ as `time' the right moving modes are forever stuck in one moment of time. Curiously what we have obtained here can be viewed as an action for a chiral Boson, constructed from an infinite number of fields.
 Note that in this case there is no $\Omega$-deformation. In higher dimensions this is not the case and, as we will now show, it will allow us to reconstruct the full higher dimensional theory. 
 
\subsection{From $2n-1$ to $2n$ dimensions}

Now we want to repeat our analysis of 2-point functions but now in  higher dimensions. For simplicity we use translational invariance to put one operator at the origin:
\begin{align}
	G_{n,k}(x^-,x^i) = \langle \hat 0|\phi^{(k)}(x^-,x^i)\phi^{(-k)}(0,0)|\hat 0\rangle \ ,
\end{align}
where $|\hat 0\rangle $ is a state in the $(2n-1)$-dimensional theory that  we identify with the $2n$-dimensional vacuum. This need   not correspond to the conventional choice of the  $(2n-1)$-dimensional vacuum but we take it to be  invariant under the $SU(1,n)$ symmetry.
Assuming spherical symmetry about the origin, we see from the action (\ref{DDscalar}) that 
 \begin{align}
\left[-\frac{2ik}{R}\partial_- +	\frac{|\vec x|^2}{4R^2}\partial_-^2    +\partial_i\partial_i\right]
G_{n,k}(x^-,x^i) = \frac{ig^2}{ \pi R}\delta(x^-)\delta^{2n-2}(x^i)\ .
\end{align}
To this end, for spherically symmetric solutions, it is helpful to introduce 
\begin{align}
z= x^- + \frac{i}{4R}x^ix^i	\ ,
\end{align}
so that the equation reduces to  
\begin{align}\label{Geq}
\left[	(z-\bar z)\partial\bar\partial  + \left(k-\frac{n-1}{2}\right)\partial+ \left({k}+\frac{n-1}{2}\right)\bar \partial \right]G_{n,k}(z,\bar z)=&-\frac{g^2}{2\pi  }\frac{i/R}{{\rm Vol}(S^{2n-3})}\left(\frac{i/2R}{z-\bar z}\right)^{n-2}\nonumber\\
&  \times\delta(z+\bar z)\delta(z-\bar z)	 \ .
\end{align}
 Ignoring the singularities at $z=\bar z=0$ we find that the  solutions are
\begin{align}\label{Gdef}
	G_{n,k}(z,\bar z) = d_{n,k}\left(\frac{1}{z\bar z}\right)^{\frac{n-1}{2}}\left(\frac{\bar z}{ z}\right)^k\Theta(x^-)\ ,
\end{align}
for some constants $d_{n,k}$. For $n=3$ this agrees with the general form for a 2-point function  in a five-dimensional theory with  $SU(1,3)$ symmetry  as constructed in \cite{Lambert:2020zdc}.

We can now reconstruct the $2n$-dimensional two-point function: 
\begin{align}
\langle\hat 0|\hat \phi(\hat x )\hat \phi(\hat 0)|\hat 0\rangle & = \sum_{k } e^{ikx^+/R}\cos^{n-1}(x^+/2R)	 \langle0|\phi^{(k)}( x^-,x^i) \phi^{(-k)}( 0,0)|0\rangle \nonumber\\
& = \cos^{n-1}(x^+/2R) 	\left(\frac{1}{z\bar z}\right)^{\frac{n-1}{2}}\sum_{k }	d_{n,k}q^k\ \Theta(x^-)\ ,
\end{align}
where
\begin{align}
	q = \frac{\bar z}{ z}e^{ix^+/R} \ .
\end{align}
Here we again encounter the problem that the sum over all $k$ will not be well-defined as $|q|=1$ and introducing an $i\varepsilon$ prescription can only cure the convergence for large $k$ or large  $-k$ but not both. 
To continue we require that positive modes Fourier modes of $\hat \phi$ annihilate the $2n$-dimensional  vacuum $|\hat 0\rangle$:
\begin{align}
	\hat\phi^{(k)}( 0,0)|\hat 0\rangle= 0\qquad k>0\ ,
\end{align}
which ensures that $\langle\hat  0| \hat\phi |\hat 0\rangle$ is invariant under translation in $x^+$.
In terms of $\phi$ this corresponds to 
\begin{align}\label{bound}
	\phi^{(k)}( 0,0)|\hat 0\rangle= 0\qquad k>-\frac{n-1}{2}\ .
\end{align}

Note that we encounter a problem if we quantize the theory using the action (\ref{DDscalar}) with $x^-$ as `time' since we obtain the conjugate momentum
\begin{align}
\Pi^{(k)}(x^-,x^i) = -\frac{2i k}{R}\phi^{(-k)}	(x^-,x^i)-\frac{|x|^2}{2R^2}\partial_-\phi^{(-k)}\ .
\end{align}
Thus $[\phi^{(k)}	(x^-,x^i),\Pi^{(k)}	(x^-,0)]=-2ikR^{-1}[\phi^{(k)}	(x^-,x^i),\phi^{(-k)}	(x^-,0)]$ is non-zero for $k\ne 0 $ and therefore we can't simultaneously impose
\begin{align}
	\phi^{(k)}( 0,0)|\hat 0\rangle=0\qquad {\rm and}\qquad \phi^{(-k)}( 0,0)|\hat 0\rangle= 0\ ,
\end{align}
which is potentially in contradiction with  (\ref{bound}).

Let us look at this more closely  on a case-by-case basis. For $n=1$ there is no problem as only positive values of $k$ appear in (\ref{bound}). For $n=2$ we must take $k$ to be half-integer so the smallest positive oscillator is $\phi^{(1/2)}$ and the bound in (\ref{bound}) becomes $k> -1/2$ which also does not include  any $\phi^{(k)}$ with $k<0$. At $n=3$ we see that we require $\phi^{(k)}|\hat 0 \rangle=0$ for $k>-1$ which includes $k=0$ along with all positive $k's$. Thus   there is no contradiction to imposing $ \phi^{(0)}|\hat 0 \rangle=0$ for $n<4$. However for $n\ge 4$ do we run into a potential problem with (\ref{bound}). 
We will not worry about this issue here as $n>3$  corresponds to CFTs in eight-dimensions or above and it is generally believed that there are no non-trivial examples. 
Thus we restrict to $n\le 3$ and are free to take $|0\rangle = |\hat 0\rangle $ with the proviso that $\phi^{(0)}|0\rangle =0$ for $n=3$. 
Note that for $n>1$, where   $\phi^{(0)}$ has a non-zero Lifshitz scaling dimension,  $\phi^{(0)}|0\rangle =0$ is also required for the vacuum to preserve $SU(1,n)$ symmetry.

To obtain the $2n$-dimensional 2-point function we need 
\begin{align}\label{dsum}
	\sum_{k \ge  (n-1)/2}	d_{n,k}q^k= C \left(\frac{i}{2R}\right)^{n-1} q^{\frac{n-1}{2}}\left(\frac{1}{1-q}\right)^{n-1}\ .
\end{align}
for some constant $C\sim g^2/\pi{\rm Vol}(S^{2n-3})$. In particular for the two cases at hand this means that must have
\begin{align}\label{dis}
d_{n,k} = C\left(\frac{i}{2R}\right)^{n-1} \left(\begin{matrix}k+\frac{n-1}{2}-1\\ k-\frac{n-1}{2}\end{matrix}\right)\ .
\end{align}
In the appendix we provide a derivation of this normalisation by requiring that we get the correct coefficient of the delta-function in (\ref{Geq}).

We also see from (\ref{dsum}) that  indeed we require $k\in {\mathbb Z}$ for odd $n$ and $k\in {\mathbb Z}+\frac12$ for $n$ even, corresponding to ensuring that $\hat \phi$ is periodic on $x^+\in[-\pi R,\pi R]$.  With these values for $d_{n.k}$ we find (again assuming an $i\varepsilon$ prescription)
\begin{align}
	\langle\hat 0|\hat \phi(\hat x )\hat \phi (\hat 0)|\hat 0\rangle &=C\cos^{n-1}(x^+/2R)\left(\frac{i}{2R}\right)^{n-1}	\left(\frac{1}{z\bar z}\right)^{\frac{n-1}{2}}\left(\frac{q^{1/2}}{1-q}\right)^{n-1}\Theta(x^-)\nonumber\\
&=C\cos^{n-1}(x^+/2R)\left(\frac{i}{2R}\right)^{n-1} \left(\frac{1}{z\bar z}\right)^{\frac{n-1}{2}}\left(\frac{1}{q^{-1/2}-q^{1/2}}\right)^{n-1}\Theta(x^-)\nonumber\\
& = C\cos^{n-1}(x^+/2R)\left(\frac{i}{2R}\right)^{n-1}\left(\frac{1}{ze^{-ix^+/2R}-{\bar z}e^{ix^+/2R}}\right)^{n-1}\Theta(x^-)\nonumber\\
&= \frac{C}{\left(-2 \hat x^+   \hat x^-  +  \hat x^i   \hat x^i\right)^{{n-1}}}\Theta(x^-)	\ .
\end{align}
Thus we recover the expected two-point function of the $2n$-dimensional theory.

\section{Conclusions and Comments}

In this paper we have examined non-Lorentzian theories with $SU(1,n)$ spacetime symmetry in ($2n-1$)-dimensions. In particular we showed how one can construct such theories by reduction of a conformally invariant Lorentzian theory in $2n$-dimensions. However other constructions may well exist. We showed  that the novel operator-state map of the Schr\"odinger group extends straightforwardly to $SU(1,n)$ theories and demonstrated how conventional non-relativistic conformal field theory is recovered in a particular limit.  We also explored some unitarity bounds and a supersymmetric extension of the spacetime symmetry algebra in five dimensions, which has been explicitly realised in a class of gauge theory examples \cite{Lambert:2019jwi,Lambert:2019fne,Lambert:2020jjm}.

We then presented examples of free theories in a variety of dimensions with various field contents. Although we kept the Kaluza-Klein tower of fields this is not necessary for $SU(1,n)$ symmetry and one can truncate the Lagrangians to a subset of Fourier modes. One can also consider including interactions ({\it e.g.} see \cite{Lambert:2019jwi,Lambert:2019fne,Lambert:2020jjm}).
We  also discussed how to reconstruct the parent $2n$-dimensional theory by keeping the entire Kaluza-Klein tower of operators. For this the role of the $\Omega$-deformation is critical.  

We note that in theories with $SU(1,n)$ symmetry we have constructed there are   terms with the `wrong-sign' kinetic term induced by the $\Omega$-deformation, when we view $x^-$ as time. However at the spatial origin such `wrong-sign' terms vanish. Given translational invariance  this suggests that the $SU(1,n)$ symmetry can be used to regain control of the theory. In particular, since there is a well-defined map to the original, non-compact, Minkowskian theory we believe that there should be a corresponding consistent treatment of the lower-dimensional theory which alleviates any such problems.

\section*{Acknowledgements}

We would like to thank A. Lipstein and P. Richmond for initial collaboration on this work, and David Tong for helpful discussions. N.L. is a co-investigator on the STFC grant ST/T000759/1, R.M. was supported by David Tong's Simons Investigator Grant, and T.O. was supported by the STFC studentship ST/S505468/1.  

This paper is to be submitted to the Frontiers Special Edition on Non-Lorentzian Geometry and its Applications. \footnote{\href{https://www.frontiersin.org/research-topics/19214/non-lorentzian-geometry-and-its-applications}{\tt www.frontiersin.org/research-topics/19214/non-lorentzian-geometry-and-its-applications}}    

\section*{Appendix: Green's Function Normalisation}

In this appendix we want to present an argument that the normalisation $d_{n,k}$ introduced in (\ref{Gdef}), which should be chosen to ensure the correct delta-function coefficient in (\ref{Geq}), does indeed agree with the form (\ref{dis}). To do this we consider an arbitrary smooth  function $f(x^-,|\vec x|)=f(z,\bar z)$ and look at the integral
\begin{align}
I[f]=\int_D f{\cal D} G_{n,k}	\ ,
\end{align}
with $G_{n,k}$ given in (\ref{Gdef}). Here $D$ is first quadrant of the $z$-plane (corresponding to $|\vec x|\ge 0$ and $x^-\ge 0$ due to the presence of $\Theta(x^-)$) and 
\begin{align}
{\cal D} G_{n,k}	=(z-\bar z)^{n-2}\left[	(z-\bar z)\partial\bar\partial  + \left(k-\frac{n-1}{2}\right)\partial+ \left({k}+\frac{n-1}{2}\right)\bar \partial \right]G_{n,k}\ .
\end{align}
We therefore need to show that we can find coefficients $d_{n,k}$ such that 
\begin{align}
	I[f] = -\frac{g^2}{2\pi  }\frac{i/R}{{\rm Vol}(S^{2n-3})}\left(\frac{i}{2R}\right)^{n-2}f(0)\ .
\end{align}
 To this end we observe that, away  from $z=0$,  we can write 
 \begin{align}
\frac{i}{2} {\cal D} G_{n,k} dz\wedge d\bar z =d\omega\ ,
 \end{align}
 where $\omega = \omega_{\bar z}d\bar z  +\omega_{z}dz$ with
 \begin{align}
 \omega_{\bar z} &= \frac{id_{n,k}}{2}\left[(1-\gamma)\left(k+\frac{n-1}{2}\right)  +\gamma\left(k-\frac{n-1}{2}\right)\frac{z}{\bar z}\right](z-\bar z)^{n-2}\left(\frac{\bar z}{z}\right)^k\left(\frac{1}{z\bar z}\right)^{\frac{n-1}{2}}\nonumber\\
 \omega_z& = -\frac{id_{n,k}}{2}\left[\gamma \left(k-\frac{n-1}{2}\right) +(1-\gamma)\left(k+\frac{n-1}{2}\right)\frac{\bar z}{z}\right](z-\bar z)^{n-2}\left(\frac{\bar z}{z}\right)^k\left(\frac{1}{z\bar z}\right)^{\frac{n-1}{2}}\ .
 \end{align}
Here $\gamma$ is an arbitrary constant corresponding to the freedom to add a total derivative $\omega\to \omega+d\Gamma$ with $\Gamma = \gamma(z-\bar z)^{n-2}G_{n,k}$. 
Thus we have
\begin{align}
I[f] & = \int_D fd	\omega \nonumber\\
& = \oint_{\partial D}f \omega - \int_D df\wedge\omega\ .
\end{align}
Next we switch to polar coordinates $z = re^{i\theta}$ and observe that
\begin{align}
\omega &= \omega_\theta d\theta	\nonumber\\
\omega_\theta &= -d_{n,k}\left[\gamma \left(k-\frac{n-1}{2}\right)e^{i\theta}+(1-\gamma) \left(k+\frac{n-1}{2}\right)e^{-i\theta}\right](e^{i\theta}-e^{-i\theta})^{n-2}e^{-2ik\theta}\ .
\end{align}
Thus if we consider $D$ as a wedge ranging between $0$ and $\pi/2$ and $r\in[0,\infty)$ then only the arc portions of the boundary  contribute to $I[f]$ so  we find
\begin{align}
I[f] 
& = f(\infty)\int_{0}^{\pi/2} \omega_\theta d\theta - \int \partial_r f \omega_\theta dr\wedge d\theta\nonumber\\
& = f(\infty)\int_{0}^{\pi/2} \omega_\theta d\theta - (f(\infty)-f(0)) \int_{0}^{\pi/2} \omega_\theta  d\theta\nonumber\\
& = f(0)\int_{0}^{\pi/2} \omega_\theta  d\theta\ .
\end{align}
To compute this integral we observe that $\omega_\theta = \partial_\theta \varphi$ with
\begin{align}
\varphi =\frac{d_{n,k}}{2i} \sum_{l=0}^{n-2}(-1)^l\left(\begin{array}{c}n-2\\ l\end{array}\right)	\left[\frac{\gamma \left(k-\frac{n-1}{2}\right)}{k+l-\frac{n-1}{2}}e^{-2i(k+l-\frac{n-1}{2})\theta}+\frac{(1-\gamma) \left(k+\frac{n-1}{2}\right)}{k+l+1-\frac{n-1}{2}}e^{-2i(k+l+1-\frac{n-1}{2})\theta}\right]\ ,
\end{align}
and hence
\begin{align}\label{answer}
\int_{0}^{\pi/2} \omega_\theta  d\theta	= \varphi(\pi/2)-\varphi(0)\ .
\end{align}
Note that if $k$ is in the range $|k|>  (n-1)/2$ then there is no value of $l$ such that the denominators in $\varphi$ vanish. For $|k| \le  (n-1)/2$ we must be more careful however, as discussed above,  we are not interested in this case here.

The integral (\ref{answer}) depends on $\gamma$ and yet $\gamma$  should not affect the Green's function $G_{k,n}$. In fact we find that $\varphi(0)$ does not depend on $\gamma$ but $\varphi(\pi/2)$ does. Thus we need to impose a   condition at $\theta=\pi/2$ (corresponding to $x^-=0$ for any $\vec x$)  where the step-function $\Theta(x^-)$ cuts off the integral. A natural choice is $\varphi(\pi/2)=0$.
This in turn fixes $\gamma$ (although its actual value is inconsequential). Proceeding in this way leads to
\begin{align}
	\int_{0}^{\pi/2} \omega_\theta  d\theta	= -\varphi(0)= -\frac{d_{n,k}}{2i}\left(\begin{matrix}k+\frac{n-1}{2}-1\\ k-\frac{n-1}{2}\end{matrix}\right)^{-1}\ .
\end{align}
Thus as a result we must take
\begin{align}
	d_{n,k}= \frac{2ig^2}{\pi  }{{\rm Vol}(S^{2n-3})}\left(\frac{i}{2R}\right)^{n-1} \left(\begin{matrix}k+\frac{n-1}{2}-1\\ k-\frac{n-1}{2}\end{matrix}\right)\ ,
\end{align}
which  reproduces (\ref{dis}).

\bibliographystyle{JHEP}
\bibliography{2112.14860}

\end{document}